\documentstyle[preprint,tighten,aps,floats,psfig,epsfig]{revtex}

\def\ul{\underline}
\def\t{\tau}

\def\b{\beta}

\def\e{\epsilon}

\newcommand{\be}{\begin{equation}}
\newcommand{\ee}{\end{equation}}
\newcommand{\bea}{\begin{eqnarray}}
\newcommand{\eea}{\end{eqnarray}}

\begin{document}

\draft
\title{Five-loop $\e$ expansion for $O(n)\times O(m)$ spin models}

\author{
Pasquale Calabrese${}^{1}$ and  
Pietro Parruccini${}^2$}  

\address{$^1$Scuola Normale Superiore and INFN,
Piazza dei Cavalieri 7, I-56126 Pisa, Italy.}

\address{$^{2}$Dipartimento di Fisica dell' Universit\`a di 
Pisa and INFN, Via Buonarroti 2, I-56100 Pisa, Italy. }

\date{\today}

\maketitle

\begin{abstract}

We compute the Renormalization Group functions of a Landau-Ginzburg-Wilson 
Hamiltonian with O$(n)\times$O$(m)$ symmetry up to five-loop in Minimal 
Subtraction scheme.
The line $n^+(m,d)$, which limits the region of second-order phase transition,
is reconstructed in the framework of the $\e=4-d$ expansion for generic 
values of $m$ up to $O(\e^5)$.
For the physically interesting case of noncollinear but planar orderings 
$(m=2)$ we obtain $n^+(2,3)=6.1(6)$ by exploiting different resummation 
procedures.
We substantiate this results re-analyzing six-loop fixed dimension series
with pseudo-$\e$ expansion, obtaining $n^+(2,3)=6.22(12)$. 
We also provide predictions for the critical exponents characterizing the 
second-order phase transition occurring for $n>n^+$.
\end{abstract}



\section{INTRODUCTION}

In the last years much effort has been dedicated in the qualitative 
understanding and quantitative description of frustrated spin systems with 
noncollinear order. Despite the intensive theoretical and experimental 
investigation, strongly debated issues remain open. In three dimensions it is 
not clear if systems as stacked triangular antiferromagnets (STAs) and 
Helimagnets should undergo a continuous phase transitions (named chiral) 
belonging to an unusual universality class and hence characterized by new 
values of critical exponents, or if the transition is first order 
(see Refs. \cite{Kawamura-98,cp-97,rev-01,DMT-03} as review).

Field-theoretical studies of such systems are based on the 
O$(n) \times $O$(m)$ symmetric Landau-Ginzburg-Wilson (LGW) 
Hamiltonian~\cite{Ham,Kawamura-98}
\begin{eqnarray}
{\cal H} = \int d^d x  \left\{ {1\over2}
      \sum_{a} \left[ (\partial_\mu \phi_{a})^2 + r \phi_{a}^2 \right]
+ {1\over 4!}u_0 \left( \sum_a \phi_a^2\right)^2 
 + {1\over 4!}  v_0
\sum_{a,b} \left[ ( \phi_a \cdot \phi_b)^2 - \phi_a^2\phi_b^2\right]
             \right\},
\label{LGWH}
\end{eqnarray}
where $\phi_a$ ($1\leq a\leq m$) are $m$ sets of $n$-component vectors.
Whether $m\leq n$ and $v_0>0$, Eq. (\ref{LGWH}) describes $m$-dimensional spin
orderings in isotropic $n$-spin space with the O$(n) \rightarrow $O$(n-m)$ 
pattern of symmetry breaking.
Negative values of $v_0$ correspond to simple ferromagnetic or 
antiferromagnetic ordering, and to magnets with sinusoidal
spin structures~\cite{Kawamura-98}. 
The condition $0< v_0< m/(m-1)u_0$ is required to have
noncollinearity and the boundedness of the free energy. 
For $m=2$ the Hamiltonian (\ref{LGWH}) describes magnets with noncollinear but
planar orderings as frustrated $XY$ ($n=2$) and Heisenberg ($n=3$)  
antiferromagnets, while for $m \geq 3$ systems with noncoplanar ground 
states~\cite{Ka-90,Kawamura-98}.

The RG flow has always two fixed points (FP's), the Gaussian one $(u_0=v_0=0)$
and the $O(n m)$  one $(v_0=0)$, while other two FP's (called chiral and
antichiral) appear for some values of $n$. 
At fixed $m$, for the physically relevant case $m \geq 1$, the $\e$ expansion 
predicts four different regimes characterized as follows \cite{Kawamura-88}:
\begin{itemize}
\item[1)] For $n>n^+(m,d)$, there are four FP's, and the chiral one 
is  stable.
\item[2)] For $n^-(m,d) < n < n^+(m,d)$, only the Gaussian and the Heisenberg
O($n m$)-symmetric FP's are present, and none of them is stable.
Thus the system is expected to undergo a first-order phase transition.
\item[3)] For $n_H(m,d) < n < n^-(m,d)$, there are again four FP's.
However for $m<7$ the stable FP is in the region $v<0$ and the transition for 
those systems with $v_0>0$ is again first order. 
\item[4)] For $n < n_H(m,d)$, \footnote{
The precise value of $n_H(m,d)$ may be inferred from the results of  
Refs. \cite{cpv-02m,cpv-03r}, where it was shown that at a global
$O(N)$ FP (here $N=nm$) all the spin four operators (and the $v_0$ term is 
spin four) have the same scaling dimension. 
Thus the $O(n m)$ FP is stable for $n<N_c/m=n_H$, where $N_c$ is the marginal 
spin dimensionality of the cubic model that results in $N_c\simeq 2.9$
in three dimensions. See the review \cite{rev-01} for an updated list of 
estimates and references about $N_c$.}
there are again four FP's, and the Heisenberg O$(nm)$-symmetric one is  stable.
\end{itemize}

For $m\geq3$ all field theoretical (FT) methods ($\e=4-d$ 
\cite{Ka-90,prv-01n}, $1/n$ \cite{prv-01n,g-02n}, fixed dimensional $d=3$ 
\cite{p-03}, $\tilde{\epsilon}=d-2$\cite{adj-90} expansions, 
and the Effective Average Action Method (EAAM) \cite{Tiss1}) agree 
indicating the existence of a marginal number of  components of the order 
parameter $n^+(m,3)$ above which a second-order phase transition is driven by 
a stable chiral FP. 

For $m=2$, again  all the above mentioned FT methods indicate the 
existence of $n^+(2)\sim 6$ above which we expect a continuous phase transition
\cite{asv-95,prv-01n,g-02n,adj-90,AS-94,lsd-00,prv-01p,cps-02,cps-03,Tiss2,Tiss3,Tiss4}. 
Instead, for $n<n^+$ the situation is controversial.
Several magnetic materials that are supposed to be described by the Hamiltonian
(\ref{LGWH}) with $n=2,3$ display scaling laws, but with experimental 
measured exponents that apparently depend on the material and sometimes 
also from the measure (see e.g. the recent experimental works \cite{exp}
and also Refs. \cite{rev-01,DMT-03} for an exhaustive list of experiments). 
The same scenario is found in MC simulations: 
systems like STA display scaling with almost well defined exponents 
\cite{Mc1} (see also \cite{Mc2} as the only simulation where apparently there
is a first-order transition in STA), 
whereas systems with the same symmetry like rigid STA (STAR) and Stiefel
$V_{n,2}$  (see e.g. Ref. \cite{DMT-03} for the definition of 
these systems) clearly undergo first-order transition for $n=2,3$ \cite{MCV}.
Unfortunately a lot of theoretical efforts have not yet produced a
unified picture.
EAAM does not provide any stable FP favoring a weak first-order phase 
transition with nonuniversal pseudo-critical 
exponents \cite{Tiss2,Tiss3,Tiss4,DMT-03} that are in rough
agreement with those coming from scaling relation found in experiments and 
Monte Carlo simulations.
Whereas in six-loop fixed dimension approach another marginal number 
of components of the order parameter $N_{c2}<n^+$ exists \cite{cps-03}, 
so that for $N_{c2}<n<n^+$ there is no stable FP, while for $n<N_{c2}$
a stable (focus-like) chiral FP appears~\cite{prv-01p,cps-02}.
Since $N_{c2}=5.7(3)$ \cite{cps-03}, according to this scenario, $XY$ and 
Heisenberg frustrated antiferromagnets undergo a second-order phase 
transition, if they are in the domain of attraction of the stable FP. 
The found critical exponents \cite{prv-01p,prv-02} are in rough agreement with 
Monte Carlo and experimental results.
Anyway, the continuous phase transition at this FP is very peculiar:
scaling properties are governed over several decades of temperatures, by 
preasymptotic effective exponents, which can differ significantly from the 
asymptotic ones, explaining in this manner the apparently scattering 
experimental and numerical results \cite{cps-02}. 
Note that the existence of a stable FP does not exclude the possibility that 
some systems may undergo a first-order transition. 
Indeed, they may lie outside the attraction domain of the stable FP.
According to mean-field arguments the systems undergoing second-order 
phase transition are those characterized by $v_0/u_0< m/(m-1)$. 
Since RG iterations may only narrow this region, systems that 
are outside this region surely undergo a first-order phase transition.


The stable FP found in fixed dimension for $n<N_{c2}$ is unrelated to the 
small-$\epsilon$ and large-$n$ chiral one.\footnote{
This statement may seem at first sight far to be trivial. 
Anyway it is quite simple to understand for $n=m=2$. In this case the $\e$
expansion has four FP's, but those with $v^*\neq0$ are in the negative half
plane~(they are the counterpart of the tetragonal model according to the 
mapping of Ref. \cite{AS-94}).
Since in $\e$ expansion a theory with $n$ couplings has at most
$2^n$ FP's, and the $O(2)\times O(2)$ has already four FP's no other FP may exist
in $\e$ expansion. Thus for $n=m=2$ it is impossible in $\e$ expansion to find
a signature of the three-dimensional chiral FP of Refs. \cite{prv-01p,cps-02}.
For $n=3$ and $m=2$ the situation is different since there are two real FP's 
and two complex zeros of $\beta$s. Anyway the latter have almost vanishing 
$v$ coordinate, and they are far from the three-dimensional chiral FP, 
ruling out the idea that there is some connection between them.
}
This fact does not imply an inconsistency of all perturbative results, 
since $\epsilon$ and $1/n$ expansions describe adiabatic moving from 4 
dimensions and $n=\infty$ respectively and they are probably inadequate to 
describe the essentially three-dimensional features of the chiral FP for 
physical systems.

\begin{table}[!t]
\begin{center}
\caption{\small Values of $n^+(m,3)$ for $m=2,3,4,5$ obtained with different
FT methods.}
\begin{tabular}{llcccc}
&Method&m=2&m=3&m=4&m=5\\
\hline
Ref. \cite{zu-94}$_{1994}$ &Local Potential Approximation &$\sim 4.7$\\
Ref. \cite{AS-94}$_{1994}$  &$d=3$ expansion: $O(g^4)$   & 3.91(1)\\
Ref. \cite{asv-95}$_{1995}$ & $\e$ expansion: $O(\e^3)$  & 3.39\\
Ref. \cite{Tiss2}$_{2000}$  &EAAM & $\sim 4$ \\
Ref. \cite{Tiss4}$_{2001}$  &EAAM & $\sim 5$ \\
Ref. \cite{prv-01n}$_{2001}$&$1/n$ expansion: $O(1/n^3)$ &5.3&7.3&9.2&11.1\\
Ref. \cite{prv-01n}$_{2001}$& $\e$ expansion: $O(\e^3)$  &5.3(2) &9.1(9) &12(1)\\
Ref. \cite{g-02n}$_{2002}$ &$1/n$ expansion: $O(1/n^2)$ &$\sim 3.24$\\
Ref. \cite{cps-03}$_{2003}$ &$d=3$ expansion: $O(g^7)$&6.4(4)&11.1(6)&14.7(8)&18(1)\\
This work & $\e$ expansion: $O(\e^5)$ &6.1(6)&9.5(5)&12.7(7)&15.7(1.0)\\
This work &pseudo-$\e$ expansion: $O(\tau^6)$ &6.22(12)& 9.9(3)&13.2(6)& 16.3(1.3)\\
\end{tabular}
\label{tabn+lit}
\end{center}
\end{table}

In recent years many efforts have been made to obtain a precise determination
of $n^+(3,m)$. A complete list of results may be found in Table \ref{tabn+lit}.
These scattered results clearly indicate that the extrapolation of 
low order $1/n$ and $\epsilon$ expansion predictions up to physical values 
of $n$ and $\epsilon$ is a quite delicate matter.
For this reason we extend the knowledge of the RG function up to five-loop 
in $\epsilon$ expansion. 
Furthermore, to improve the goodness of fixed-dimensional $d=3$ estimates we 
re-analyze the six-loop perturbative series~\cite{prv-01p,p-03} with the 
pseudo-$\e$ expansion trick~\cite{Ni}, since
in many cases this method provided the most accurate results in the 
determination of the marginal number of order parameter components (see Ref. 
\cite{FHY-00} for the cubic model and Ref. \cite{HDY-01} for the weakly 
diluted $m$-vector model). We anticipate that also in this case the final 
estimate is very precise, as clear from Table \ref{tabn+lit}.

The paper is organized as follows. In Section \ref{sec2} we determine 
$n^\pm(m,3)$ for several $n$ by means of five-loop $\e$ expansion.
Pseudo-$\e$ expansion is considered in  Section \ref{sec3}. 
Section \ref{concl} summarizes our results.
In the Appendix \ref{Appexp} we report the estimates of the exponents 
for $m=2$ and $n\geq n^+(2,3)$, whereas in the Appendix \ref{appA}  
details of series computation are presented, and the Appendix \ref{appB} 
reports analytic forms of some quantities given in the text.

\section{Five-loop $\epsilon$-expansion}  
\label{sec2}

We extend the three-loop $\epsilon$ expansion of Refs. \cite{asv-95,prv-01n}
for the RG functions of the O($n$)$\times$O($m$) symmetric theory to five-loop.
The obtained series and some details of calculations are reported in the
Appendix \ref{appA}.
Within these series, the $\epsilon$ expansion of 
$n^\pm(m,4-\e)$ may be calculated to $O(\epsilon^5)$. 
They are expanded as
\be
n^\pm(m) = n_0^\pm(m) + n_1^\pm(m)\, \epsilon + n_2^\pm(m)\, \epsilon^2  
 +n_3^{\pm}(m)\, \e^3+n_4^{\pm}(m)\, \e^4   +      O(\epsilon^5),
\ee
and the coefficients $n_i^\pm(m)$ may be obtained by requiring 
\be
\beta_u(u^*,v^*;n^\pm) = 0, \qquad \qquad
\beta_v(u^*,v^*;n^\pm) = 0, 
\ee
and 
\be 
{\rm det}\, \left|{\partial(\beta_u,\beta_v)\over \partial(u,v)} \right| 
 (u^*,v^*;n^\pm) = 0.
\ee
Note that the previous equation is not the only possible choice to 
find $n^\pm$, one can e.g. impose the coincidence of the coordinates
of the chiral and antichiral FP etc. However this was the most advantageous 
from the computational point of view.  
For generic values of $m$, the results obtained for the location of the 
FP's and for $n^\pm(m)$ are too cumbersome in order 
to be reported here, thus we  report them only at fixed $m=2,\,3$.

\subsection{Noncollinear but planar ordering: m=2}

Let us first consider the physically relevant case $m=2$.
The numerical expression for $n^\pm(2,4-\e)$ simplifies to
(their analytical expressions are in App. \ref{appB})
\begin{eqnarray}
n^+(2,4-\e)&=& 21.7980-23.4310 \,\e + 7.0882\,\e^2-0.0321 \,\e^3
+4.2650\, \e^4+O(\e^5)\,,
\label{n2+}\\
n^-(2,4-\e)&=& 2.2020-0.5690 \,\e + 0.9892\,\e^2-2.2786 \,\e^3+6.5406\, \e^4+O(\e^5)\,\nonumber.
\end{eqnarray}
The first three terms of the expansions are in agreement with previous 
works \cite{asv-95,prv-01n}.
In order to give a numerical estimate of $n^{\pm}(2,3)$ such series should be
evaluated at $\e=1$. 
A direct sum is obviously not effective since the series are clearly divergent.
The series $n^-(2,4-\e)$ has a quite regular behavior in $\e$, with the last 
terms approximately factorial growing and with alternating sign coefficients.
Thus a Pad\'e-Borel-Leroy (PBL) resummation~(see App. \ref{appA}) should be 
effective, and in fact it provides $n^-(2,3)=1.968(1)$.
Despite to the stability of the result, this estimate
cannot be right. Indeed we known, from the mapping onto the 
tetragonal model \cite{AS-94}, that for $n=m=2$ a couple of
FP's exists in the $\e$ expansion, thus $n^-(2,3)\geq 2$ holds.
A direct analysis of three-dimensional six-loop series shows that $n^-(2,3)$ is
a bit larger than 2 \cite{unp}.
This was firstly noted in Ref. \cite{AS-94}.
Anyway the precise value of $n^-(2,3)$ is not of interest for frustrated 
models and this disagreement will be not more commented below. 
We only quote it to show that PBL method apparently underestimates the 
final error bar.

\begin{table}[!tbp]
\caption{Non-defective PBL results for $n^+(m=2,3)$. The estimate from these data is $5.47(7)$.}
\label{tab1}
\begin{center}\begin{tabular}{l|c|cc|cc}
\multicolumn{1}{l|}{Loop}&
\multicolumn{1}{c|}{3}&
\multicolumn{2}{c|}{4}&
\multicolumn{2}{c}{5}\\
Pad\'e &[1/1]&[1/2]&[2/1]&[1/3]&[3/1]\\
\hline\hline
b=0&3.385& 4.477&  5.423  &5.402&5.455\\
b=1&3.514& 4.593&  5.423  &5.586&5.455\\
b=2&3.583 &4.654&  5.423  &5.688&5.455\\
b=3&3.626 &4.693&  5.423  &5.753&5.455\\
\end{tabular}\end{center}\end{table}

The evaluation of  $n^+(2,3)$ is much more difficult because of its 
irregular behavior.
The results of PBL resummations are shown in Table \ref{tab1} for several
approximants. The fifth order ones indicate $n^+(2,3)= 5.47(7)$~(it is 
the average and the variance of the approximants  [1/3] and [3/1]
with $b=0,1$). Anyway the previous analysis of $n^-(2,3)$ suggests that 
PBL method underestimates error bars, thus 
to have a more reliable result, we  apply other summation 
methods to evaluate $n^+(2,3)$, that in practice allow one to extract better 
behaved series from Eq. (\ref{n2+}). 
In what follows we present three different functions of $n^+(2,3)$ that behave
better than Eq. (\ref{n2+}). The choice of the functions is motivated or 
by physical reasons (as Eq. (\ref{cons})) or by the experience in resumming 
divergent series (as Eq. (\ref{1/n})). Obviously we do not exclude the 
possibility of better choices.

The first function (as usually done for critical exponents) is
\be 
{1 \over n^+(2,4-\e)}=0.0459 + 0.0493\e + 0.0381\e^2  + 0.0250\e^3 +0.0056\e^4+O(\e^5)\,,
\label{1/n}
\ee
whose coefficients decrease rapidly.
Setting $\e=1$, without resummation,\footnote{
We do not use any resummation since the series is convergent up to the 
considered order. A good resummation should reproduce the result of the direct
sum. Anyway note that PBL is not expected to work, since the series
has not alternating signs. The same remark holds for Eq. (\ref{1/a}).
}
one obtains   
$n^+(2,3)=7.50$ (at 3-loop), $n^+(2,3)=6.31$ (at 4-loop), and 
$n^+(2,3)=6.10$ (at 5-loop).

Another method, firstly employed in Ref.~\cite{prv-01n}, use the knowledge
of $n^+(2,2)$ to constrain the analysis at $\e=2$, under the assumption that 
${n^+}(m,d)$ is  sufficiently smooth in $d$ at fixed $m$.
In Ref.~\cite{prv-01n} it is assumed that the two-dimensional LGW stable 
FP is equivalent to that of the NL$\sigma$ model for all $n\ge 2$, 
$m\ge 2$, except $n=2$, $m=2$. Since  the NL$\sigma$ model is asymptotically
free, the authors of Ref. \cite{prv-01n} conclude that $n^+(2,2)=2$.
The knowledge of ${n}^+(2,2)$ may be exploited in order to 
obtain some informations on $n^+(2,3)$, rewriting $n^+(2,4-\e)$ as 
\bea
n^+(2,4-\e)&=&2 + (2 - \e) (9.8990 - 6.7660 \e +0.1611\e^2 +  0.0645 \e^3  + 2.1648 \e^4 )+O(\e^5)\nonumber \\
&=&2 + (2 - \e)\,a(m=2,\e).
\label{cons}
\eea
Since the coefficients of $a(2,\e)$ do not decrease, we
consider $1/a(2,\e)$ obtaining the more 
``convergent'' expression
\be
{1 \over a(2,\e)}=0.1010 + 0.0690 \e + 0.0456 \e^2  + 0.0294 \e^3  - 0.0032\e^4+O(\e^5),\label{1/a}
\ee 
which, setting $\e=1$, gives
$n^+(2,3)=6.64$ at three loop, $n^+(2,3)=6.08$ at four loop, and 
$n^+(2,3)=6.136$ at five loop.

Actually the exact value of $n^+(2,2)$ is a controversial matter, in fact
it was pointed out that $Z_2$-type topological defects may lead to a 
finite-temperature phase transition in the two-dimensional $O(3)\times O(2)$ 
model (see e.g. Refs. \cite{cp-01,french2d} and references therein, for 
different scenarios about the critical behavior of this model).
Independently from the fact that the perturbative LGW 
approach is able to describe such topological excitations (that is still a very
controversial point), one should expect as naive upper bound $n^+(2,2) = 4$, 
since for $n\geq 4$ the NL$\sigma$ surely works. 
Using as constraint first $n^+(2,2)=2$ and then $n^+(2,2)=4$ one
should kept in between the right value.
Taking $n^+(2,2)=4$ and considering again the 
series of the inverse of the resulting $a(2,\epsilon)$, we have 
$n^+(2,3)=7.57$ at three loop, $n^+(2,3)=6.91$ at four loop, and 
$n^+(2,3)=6.70$ at five loop.
The five-loop results with imposing $n^+(2,2)=2$ and 
$n^+(2,2)=4$ are quite close, signaling a weak dependence from the 
exact value of $n^+(2,2)$ and making the constrained analysis quite safe.

\begin{figure}[t]
\centerline{\epsfig{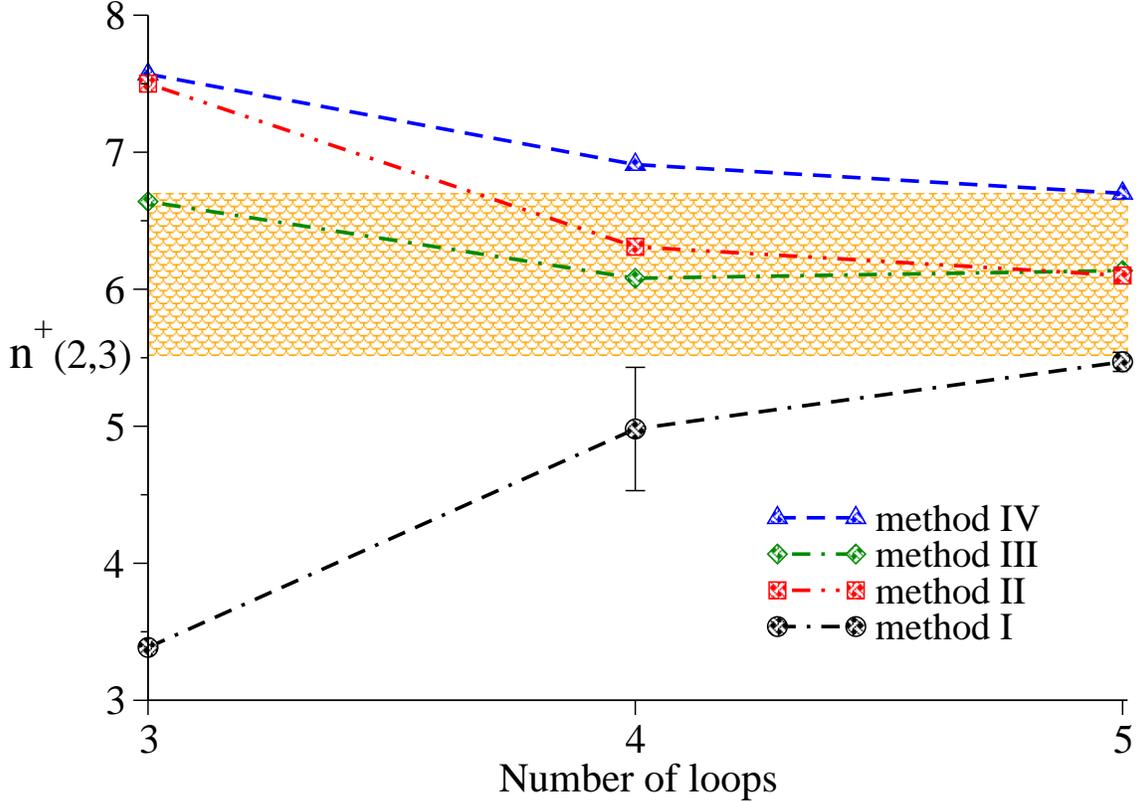}}
\caption{Different estimates of $n^+(2,3)$ with varying the number of loops.
Method I is PBL, method II is Eq. \ref{1/n}, method III is Eq. \ref{1/a}, and
method IV is the same of III, but with $n^+(2,2)=4$. The shadow region is our
 final estimate $n^+(2,3)=6.1(6)$.}
\label{fig}
\end{figure}

In order to give a final estimate with a proper error bar, we report in Fig. 
\ref{fig} all our results.
It appears that all the employed summation methods are {\em slowly} converging 
to the same asymptotic value, in particular
PBL converges from below, while the other three from above and quickly.
Note in particular that Eq. (\ref{1/a}) (method II in figure) is quite flat
up to the considered order, suggesting a good convergence. 
We take as final estimate the very conservative one $n^+(2,3)=6.1(6)$, 
which includes all five-loop data.
The PBL result is only marginally compatible with the final estimate, but 
this is not surprising since PBL is the method that converges slowly and it 
is expected to underestimate error bars from the analysis of $n^-(2,3)$ 
(note also that the 
three-loop PBL 3.39 is the almost half of the common accepted value).
The final value is only marginally compatible with the three-loop 
estimate $5.3(2)$ of Ref. \cite{prv-01n}. Furthermore our quoted error is
three times bigger than the three-loop one. We believe that our estimate is 
safe, contrarily to Ref. \cite{prv-01n} were probably the 
valuation of the error bar was too optimistic.

The problem of the value of $n^+(2,2)$ may be also exploited resumming the 
series for $n^+$ at $\epsilon=2$, assuming a smooth dependence
on $d$ up to $\epsilon=2$. 
The behavior of the series is worst than that for $\epsilon=1$, in fact 
Eqs. (\ref{1/n}) and (\ref{cons}) are not effective anymore.
However we can obtain useful informations from PBL method.
The approximants [1/1], [1/2], and [1/3] give negative values and must 
be discarded. Only two working approximants remain, namely
the [2/1] which leads to $n^+(2,2)=3.0$ and the [3/1] providing $n^+(2,2)=3.3$,
independently from $b$. As a consequence we have a weak indication for
$n^+(2,2)$ greater than the conjectured value $2$.


\subsection{Nonplanar ordering: m=3}

Let us now consider the case $m=3$ which is relevant for spin systems which 
possess nonplanar orderings of the ground state. 
One possible example of such noncoplanar criticality was supposed to be 
the pyrochlore antiferromagnets FeF$_3$ \cite{Pyr} and \cite{Kawamura-98} 
(page 4733).
The coefficients $n_i^{\pm}(3)$ are (the analytic expressions are in the 
Appendix \ref{appB})
\bea 
n^+(3,4-\e)&=&32.4919-33.7184 \e + 11.1002 \e^2  - 2.1440 \e^3  + 5.2756\e^4+O(\e^5),\label{n3+}\\
n^-(3,4-\e)&=&1.5081-0.2816\e+0.5827\e^2  -1.4192\e^3+4.0193\e^4+O(\e^5).\nonumber
\eea
Again for $\epsilon=1$, PBL technique for $n^-(3,3)$ seems very stable, 
leading to  $n^-(3,3)= 1.409(1)$. 
For $n^+(3,3)$ we can repeat the same  analyses used for $m=2$.
The PBL results are reported in Tab. \ref{tab2}.
In this case they are much more sensitive to the used approximant and a 
reliable estimate is difficult. Anyway this analysis suggests $n^+\sim 9$.

\begin{table}[!tbp]
\caption{Non-defective PBL results for $n^+(m=3,3)$.}
\label{tab2}
\begin{center}\begin{tabular}{l|c|cc|cc}
\multicolumn{1}{l|}{Loop}&
\multicolumn{1}{c|}{3}&
\multicolumn{2}{c|}{4}&
\multicolumn{2}{c}{5}\\
Pad\'e &[1/1]&[1/2]&[2/1]&[1/3]&[3/1]\\
\hline\hline
b=0&6.448& 7.591 &8.152& 8.621&  9.176\\
b=1&6.654& 7.705 &8.134& 8.834&  9.191\\
b=2&6.763& 7.760 &8.123& 8.954&  9.202\\
b=3&6.831& 7.792 &8.116& 9.031&  9.209\\
\end{tabular}\end{center}\end{table}

As for $m=2$, the situation is better with considering 
\be
{1 \over n^+(3,4-\e)}= 
0.0308 + 0.0319\,\e + 0.0226\,\e^2 + 0.01460\,\e^3 + 0.0045\,\e^4+O(\e^5),
\ee
which gives rise to $n^+(3,3)=11.7$ at three-loop, $n^+(3,3)=10.0$ at four-loop
and $n^+(3,3)=9.57$ at five-loop.

Following Ref. \cite{prv-01n} we impose in the constrained analysis only 
$n^+(3,2)=2$, since (at least for what we known) this value was never 
criticized in literature. We obtain
\bea
n^+(3,4-\e)&=&2 + ( 2 -  \e) \,( 15.2460 - 9.2362\,\e + 0.9320\,\e^2 - 0.6060\,\e^3 + 2.3348\,\e^4 ) +O(\e^5)\nonumber \\
&=&2 + (2 - \e)\,a(m=3,\e)\,,
\eea
and again 
\be
{1 \over a(3,\e)}=0.0656 + 0.0397\e + 0.0201\e^2  + 0.0123\e^3-0.0022\e^4+O(\e^5),
\ee
has a better behavior, giving at $\e=1$,  $n^+(3,3)=9.975$, 
$n^+(3,3)= 9.26$, $n^+(3,3)=9.38$ at three-, four-, and five-loop respectively.

The data are much more scattered than those for $m=2$. 
A possible estimate, which includes the majority of the previously obtained 
results, is $n^+(3,3)\sim9.5(5)$, that is lower than the six-loop 
fixed-dimension prediction $n^+(3,3)=11.1(6)$ \cite{p-03}.


In order to give a complete picture of the critical behavior for all 
values of $n$ and $m$, we calculate the expansion of $n^+(m,4-\e)$ 
$m=4,5$, obtaining (applying the same summation methods of before)
$n^+(4,3)=12.7(7)$ and $n^+(5,3)=15.7(1.0)$, that are the values reported 
in Table \ref{tabn+lit}. These results and their comparison with 
fixed dimension ones will be commented in the Conclusions.

\section{Pseudo $\e$ expansion for \lowercase{ ${n^+(m,3)}$}}
\label{sec3}

In this section we re-analyze the six-loop three-dimensional series of 
Refs. \cite{prv-01p,p-03} with the 
pseudo-$\e$ expansion trick, since, as we have mentioned in the Introduction, 
it provided very good results for the marginal spin dimensionality in
other models. 

The idea behind this trick is very simple: one has only to multiply the linear 
terms of the two $\beta$ functions by a parameter $\t$, find the FP's~(i.e.
the common zeros of the $\beta$'s) as series in $\t$ and analyze the results 
as in the $\e$ expansion.
If one is interested in the critical exponents, they are obtained as series 
in $\t$ inserting the FP expansion in the appropriate RG functions. 
With this trick the cumulation of the errors coming from the non-exact 
knowledge of the FP and from the uncertainty in the resummation of the 
exponents is avoided, obtaining very precise results.
Note that now, differently from $\e$ expansion, only the value 
at $\t=1$ makes sense. 
The only drawback of this trick is that it is unable to capture physical 
features of the systems that appear at higher perturbative orders.
In fact at one-loop the series are equal (apart a normalization) to those
of the $\e$ expansion. Thus the FP's structure may not be different from that
of $\e$ expansion.
All the remarks and the limits of $\e$-expansions apply also to pseudo-$\e$ 
one.
Explicitly the pseudo-$\e$ expansion is able only to evaluate $n^\pm(m,3)$ and 
$n_H(m,3)$ but not $N_{c2}$. For this reason the pseudo-$\e$ expansion is
unable to recover the critical behavior for $m=2$ and $n=2,3$ coming from 
fixed dimensional field theory, even if it is a re-analysis of the same series.

For $m=2$ we obtain 
\be
n^+(2,3)=21.7980-15.6206 \,\t
+0.2621\,\t^2-0.1509\,\t^3-0.0392\,\t^4-0.0299\,\t^5+O(\t^6)\,.
\ee
At least up to the presented number of loops the series does not
behave as asymptotic ones with factorial growth of coefficients and 
alternating signs. So one may apply a simple Pad\'e resummation of the series.
The results are displayed in Tab. \ref{paden+}. Several approximants have 
poles on the positive real axis (these Pad\'e are the underlined numbers)
close to $\t=1$ and thus the estimates of $n^+$ on their basis should be 
considered unreliable.
Anyway some of these defective approximants have poles ``far'' from 
$\t=1$, where the series must be evaluated. Thus one may expect 
the presence of a pole not to influence the approximant at $\t=1$. 
Indeed all such Pad\'e results are very close since lower order.
We choose as final estimate the average of those six-loop order Pad\'e without
poles at $\t<2$~(excluding those with $N=0$, giving unreliable results), and 
as error bar we take the maximum deviation from the average of four- and 
five-loop  Pad\'e.
Within this procedure we have $n^+(2,3) = 6.22(12)$.
In order to corroborate this value we resum the series even with the PBL.
In this case only the three-loop [1,1] and the four-loop [2,1] approximants
are nondefective, giving $n^+(2,3)=6.43$ and  $n^+(2,3)=6.35$ respectively, 
in agreement with the Pad\'e estimate at the same perturbative order.
Anyway a six-loop estimate is not possible within PBL.

\begin{table}[!t]
\begin{center}
\caption{\small Pad\'e table for the series $n^+(2,3)$. The underlined values 
come out from Pad\'e which possess poles on the positive real axis.
The location of the poles nearest to $\t=1$ is reported in brackets.
The final estimate is $n^+(2,3)=6.22(12)$.}
\begin{tabular}{|l|cccccc|}
& $N=0$ & $N=1$ &$N=2$ &$N=3$ &$N=4$ &$N=5$ \\
\hline
$M=0$&21.798&6.177&6.439&6.289&6.249&6.219\\
$M=1$&12.698 &6.435 &6.344&\ul{6.236}\small{\{3.85\}}&\ul{6.123}\small{\{1.31\}}&\\
$M=2$&9.827&\ul{ 6.290}\small{\{10.38\}}&\ul{ 6.230}\small{\{3.04\}}&\ul{ 6.182}\small{\{1.74\}}&&\\
$M=3$&8.463& \ul{6.247}\small{\{5.81\}}&\ul{6.154}\small{\{1.45\}}&&&\\
$M=4$&7.695&\ul{6.217}\small{\{3.87\}}&&&&\\
$M=5$&7.221&&&&&\\
\end{tabular}
\label{paden+}
\end{center}
\end{table}

To find $n^+(m,3)$ for systems with nonplanar spin orderings, we apply
the same procedure exploited for $m=2$ obtaining the following perturbative 
expressions  up to $O(\t^6)$
\bea
n^+(3,3)&=&32.4919-22.4789\,\t
+0.4633\t^2-0.2628\t^3-0.1429\,\t^4-0.1123\t^5,
\eea
Again, to extract quantitative information from them, we apply
the procedure of before and check the results with PBL method.
The final estimates are displayed in Tab. \ref{tabn+lit} together with those 
coming from five-loop $\e$ expansion and other field theoretical methods,
also for $m=4,5$.

\section{conclusions}
\label{concl}

We compute the Renormalization Group functions of a Landau-Ginzburg-Wilson 
Hamiltonian with O$(n)\times$O$(m)$ symmetry up to five-loop in Minimal 
Subtraction scheme.
Such higher order computation allowed us to reconstruct with good precision
the line $n^+(m,d)$ in $\e$ expansion up to $O(\e^5)$.
We have also re-analyzed six-loop fixed dimension series
with pseudo-$\e$ expansion.
All the results for $n^+(m,3)$ with $m=2,3,4,5$ are reported in 
Table \ref{tabn+lit}.
The results coming from the $1/n$ expansion are not so near in magnitude to
the $\e$ and pseudo-$\e$ ones. Anyway, in this case $n^+$ is
determined by perturbative series which are known only to $O(1/n^2)$, 
the correction to the leading term is not small, and since $n^+$ is not 
expected to be large in magnitude, one is extrapolating the results to 
value of $n$ which could be dangerous.
Also the agreement with previous three-loop $\e$-expansion is not good.
Anyway, as already stated in the text, we believe that the error 
quoted in previous three loop works Refs. \cite{asv-95,prv-01n} was
underestimated.  
Differently the agreement is good between all the higher order perturbative
methods, i.e. direct six-loop fixed dimension, $\e$, and pseudo-$\e$ expansion.
Note that the pseudo-$\e$ results are between the five-loop $\e$ expansion 
and direct six-loop fixed dimension ones and are the more credible since
they are the more stable with changing the perturbative order.

For the physically interesting case of noncollinear but planar orderings 
$(m=2)$, we obtain $n^+(2,3)=6.1(6)$ from five-loop $\e$ expansion
 and $n^+(2,3)=6.22(12)$ from six-loop pseudo-$\e$ expansion.
Note that the last value seems to exclude $n=6$ from the second-order phase 
transition region, in contradiction with Monte Carlo results ~\cite{lsd-00} 
and EAAM \cite{Tiss4} . Anyway two remarks on this point are necessary.
First, being $n^+(2,3)-6=0.22(12)$  small, the concept of pseudo-scaling
discussed in Ref.~\cite{Tiss3,DMT-03} applies and the transition is expected 
to be extremely weak first order, i.e. all RG trajectories are attracted 
toward a small domain where the flow is very slow.
In such domain the system spends the majority of its RG time (making the
correlation length very large) and scaling is partially recovered.
Second, the estimate of $N_{c2}=5.7(3)$ is marginally compatible with 6, 
leaving the possibility of a second-order phase transition at $n=6$ in the
scenario of Refs. \cite{prv-01p,cps-02,cps-03}.
Anyhow we should expect measured critical exponents for $n=6$ 
close to those at $n^+(2,3)$. 
Monte Carlo \cite{lsd-00} and EAAM \cite{Tiss4} provide
$\nu=0.700(11)$, $\gamma=1.383(36)$, and 
$\nu\simeq 0.707$, $\gamma\simeq 1.377$ respectively, 
that are quite close, but definitely different from the FT perturbative 
results for $n= n^+$ $\nu_{n=n^+}=0.635(4)$ (see Appendix \ref{Appexp}). 
We believe that this apparent disagreement deserves further investigations.

\section*{ACKNOWLEDGMENTS}

We thank Andrea Pelissetto, Paolo Rossi and Ettore Vicari for
useful and interesting discussions.
We thank Yu. Holovatch, D. Ivaneyko and M. Tissier for pointing out two misprints in Tables \ref{6loopbu} and \ref{6loopbv}.

\appendix

\section{Critical exponents}
\label{Appexp}

\begin{table}[t]
\squeezetable
\caption{
Critical exponents for $n>n^+(2,3)$. Values obtained with other FT methods 
are reported for comparison. The EAAM and MC results for $n=6$ are also
reported}.
\begin{tabular}{l|c|c|c|c|c|c}
$\nu$ &$n=6$&$n=n^+$ &$n=7$&$n=8$ &$n=16$ & $n=32$   \\
\hline
This work && 0.635(4) &0.71(4) & 0.75(4) & 0.89(4) & 0.94(2) \\
6L FD\cite{cps-03}&& &0.68(2)    & 0.71(1)  & 0.863(4) & 0.936(1)   \\
O$(1/n^2)$ \cite{prv-01n}&& & 0.697   & 0.743 & 0.885 & 0.946  \\
EAAM&0.707&&&&&\\
MC&0.700(11)&&&&&\\
\tableline \hline
$\gamma$&$n=6$&$n=n^+$& $n=7$&$n=8$ &$n=16$ & $n=32$  \\
\hline
This work &&1.25(2)&1.39(6) &1.45(6)&1.75(4)&1.87(4)\\
6L FD\cite{cps-03}& & & 1.31(5) & 1.40(2) & 1.70(1) & 1.860(5) \\
$O(1/n^2)$ \cite{prv-01n}&& & 1.36   & 1.45 & 1.75 & 1.88  \\
EAAM&1.377&&&&&\\
MC&1.383(36)&&&&&\\
\tableline \hline
[$\omega_1$, $\omega_2$]&$n=6$&$n=n^+$  & $n=7$ & $n=8$ & $n=16$ & $n=32$ \\
\hline
This work &&[0,0.86(3)]&[0.84(3),0.33(10)] &[0.84(3),0.45(8)] & 
[0.86(1),0.77(2)] &[0.91(2),0.90(1)]\\
6L FD\cite{cps-03}&&& [0.83(2),0.23(5)]  & [0.83(2),0.36(4)] & 
[0.876(4),0.714(9)] & [0.933(2),0.868(2)]    \\
O$(1/n)$ \cite{g-02n}&&&
[0.768, 0.537]&[0.797,0.594]&[0.899,0.797]&[0.949,0.899] \\
\end{tabular}
\label{tabexp}
\end{table}

The critical exponents are computed by expanding in power of $\epsilon$ the 
exponent series at the FP. 
Such a computation gives the exponents only for 
\be
    n > n^+_0(m)=5 m + 2 + 2 \sqrt{6 (m+2)(m-1)},
\ee
or
\be
    n < n^-_0(m)=5 m + 2 - 2 \sqrt{6 (m+2)(m-1)}.
\ee
Indeed, if these bounds are not satisfied the FP's are complex 
and therefore also the exponent series.
This is actually only an artifact of the $\e$ expansion and the exponents are 
real and well defined for $n>n^+(m,d)$~(see for details 
Refs. \cite{Kawamura-88,prv-01n}). 
In order to obtain series for the exponents in all the relevant domain we 
can perform the following trick \cite{prv-01n}. 
For $n^+<n < n_0^+$ we set $n = n^+(m,4-\e) + \Delta n$ and re-expand all 
series in powers of $\e$ keeping $\Delta n$ fixed. 
In particular, for $\Delta n=0$ we obtain the critical exponents for $n=n^+$. 
In the most interesting case of $m=2$ the exponents along the line $n=n^+$ are
(their analytic forms are in the App. \ref{appB})
\be
\eta_{n=n^+} =0.0208\,\e^2+0.0174\,\e^3+0.0061\,\e^4+0.0032\,\e^5+O(\e^6)\,,
\label{etan+}
\ee
and
\be
(\nu_{n=n^+})^{-1} =2 - 0.5 \,\e + 0.0290 \,\e^2  + 0.0767 \,\e^3  - 0.0478 \,\e^4+0.0762 \,\e^5+O(\e^6)\,.
\label{nun+}
\ee
After a PBL resummation we have $\nu_{n=n^+}=0.635(4)$  and 
$\eta_{n=n^+}=0.049(1)$. 
In the same way we have also estimated the subleading exponents governing the 
corrections to the scaling~(i.e. the eigenvalues of the matrix (\ref{omega})). 
The smallest one is obviously $\omega_1^{n=n^+}=0$, instead the biggest one 
is $\omega_2^{n=n^+}=0.86(3)$, after PBL resummation.
Now we may calculate critical exponents for all $n>n^+(2,3)$, even if
$n<n_0^+(2)\simeq 21.8$. 
The resulting critical exponents have a double source of error: 
one coming from uncertainty in the resummation and one  from 
the not precise knowledge of $n^+(2,3)$.
Note that the latter dominates the given error for $n=7,8,16$, since the 
resummation of the exponents is very precise, contrarily to that of $n^+(2,3)$.
To give an example, at $n=7$ we have $\Delta n=0.9(6)$, that 
leads to $\nu(\Delta n=0.3)\simeq 0.67$ and $\nu(\Delta n=1.5)\simeq 0.75$,
and thus the resulting estimate $\nu(n=7)=0.71(4)$. All other exponents are
calculated in the same way.  
For $n>n_0^+(2)$ a direct resummation of the critical exponents is possible.
In Table \ref{tabexp} we report the results for $n=7,\,8,\,16,\,32$ and 
the corresponding six-loop fixed-dimension and $1/n$ ones for comparison.
There is a quite good agreement between all FT methods, a part from the 
estimates of $\omega$ at $n=16,32$, that seems to differ significantly from 
fixed dimension results. The source of such disagreement is not completely
clear to us. It may be due to an underestimating of the error (especially in 
fixed dimension), or to the bad behavior of the resummation. Anyway it is
not so surprising, since already for the $O(n)$ model the estimates of 
$\omega$ are not very good (see e.g. Ref. \cite{rev-01}).

\section{Perturbative Series}
\label{appA}

We calculate the perturbative RG function in the minimal subtraction 
($\overline{\rm MS}$) renormalization scheme for the massless theory.
We compute the divergent part of
the irreducible two-point functions of the field $\phi$,
of the two-point correlation functions with insertions of the 
quadratic operators  $\phi^2$, and 
of the two independent four-point correlation functions.
The diagrams contributing to this calculation are 162 for the four-point 
functions and 26 for the two-point one.
We handle them with a symbolic manipulation program, which  
generate the diagrams and compute the symmetry and group factors of 
each of them. We use the results of  Ref.~\cite{KS-01}, 
where the primitive divergent parts of all integrals appearing in our 
computation are reported.
We determine the renormalization constant $Z_{\phi}$ 
associated with the fields $\phi$, the renormalization constant  $Z_t$ of the 
quadratic operator $\phi^2$, and the renormalized quartic couplings $u,v$.
The functions $\beta_u$, $\beta_v$, $\eta_\phi$ and  $\eta_t$ are 
determined using the relations
\be
\beta_u (u,v) = \mu \left. {\partial u \over \partial \mu} \right|_{u_0,v_0},
\qquad\qquad
\beta_v (u,v) = \mu \left. {\partial v \over \partial \mu} \right|_{u_0,v_0},
\ee
\be
\eta_\phi (u,v)= 
  \left. {\partial \log Z_\phi \over \partial \log \mu} \right|_{u_0,v_0},
\qquad
\eta_t (u,v)= \left. {\partial \log Z_t \over \partial \log \mu} \right|_{u_0,v_0}.
\ee

The zeroes $(u^*,v^*)$ of the $\beta$ functions provide the FP's
of the theory. In the framework of the $\epsilon$ expansion, they are obtained 
as perturbative expansions 
in $\epsilon$ and then are  inserted in the RG functions to determine
the $\epsilon$ expansion of the critical exponents:
\be
\eta=\eta_\phi (u^*,v^*), \quad
\nu =(2 - \eta_\phi (u^*,v^*) + \eta_t(u^*,v^*))^{-1}.
\ee
The stability of each FP is controlled by the $2\times 2$ matrix 
\begin{equation}
\label{omega}
\Omega = \left(\matrix{\displaystyle 
\frac{\partial \beta_u(u,v)}{\partial \,u} 
&\displaystyle \frac{\partial \beta_u(u,v)}{\partial \,v}
 \cr \cr \displaystyle 
\frac{\partial \beta_v(u,v)}{\partial \,u} 
& \displaystyle  \frac{\partial \beta_v(u,v)}{\partial \,v}}\right)\; .
\end{equation}
A stable FP must have two eigenvalues with positive real 
parts, while a FP possessing eigenvalues with nonvanishing 
imaginary parts is called focus.

In reporting the series we make explicit use of the symmetry under the exchange
$(m,n)$ and write the RG functions as
\bea
\beta_u=\b_1^u+\b_2^u+\b_3^u+\b_4^u+\b_5^u+\b_6^u,&\qquad& 
\b_k^u=\sum_{i=0}^k u^iv^{k-i} (U_{k-i,i}(m,n)+U_{k-i,i}(n,m)),  \\
\beta_v=\b_1^v+\b_2^v+\b_3^v+\b_4^v+\b_5^v+\b_6^v,&\qquad& 
\b_k^v=\sum_{i=0}^k u^iv^{k-i}( V_{k-i,i}(m,n)+V_{k-i,i}(n,m)), \\
\eta_\phi=\eta_1+\eta_2+\eta_3+\eta_4+\eta_5,&\qquad& 
\eta_k=\sum_{i=0}^k u^iv^{k-i}( E_{k-i,i}(m,n)+ E_{k-i,i}(n,m)),\\
{1 \over \nu}=\nu_1+\nu_2+\nu_3+\nu_4+\nu_5,&\qquad& 
\nu_k=\sum_{i=0}^k u^iv^{k-i} (B_{k-i,i}(m,n)+B_{k-i,i}(n,m)), 
\eea
where the first three orders coincide with those of 
Refs.~\cite{Kawamura-88,Ka-90,asv-95,prv-01n}, and 
the four and five-loop coefficients are reported in
Tabs. \ref{t1}, \ref{6loopbu}, \ref{6loopbv}, \ref{tabeta}, 
\ref{tabnu}.\footnote{The complete list of the series is available in electronic format 
on request.}

In order to verify the exactness of our perturbative series we perform 
several checks on them. 
They reduce to the existing $O(\epsilon^5)$ ones 
for the O($N$)-symmetric theory \cite{ON} in the proper limit. 
For the particular case $n=m=2$ they agree, according to the exact mapping of 
Ref. \cite{AS-94}, with the four-loop $\epsilon$ expansion 
of the so called tetragonal model Ref.~\cite{MV-00}. 
The value of $n_H(m,4-\e)$ coincides with $N_c/m$, where $N_c$ is the
marginal spin dimensionality of the cubic model obtained in \cite{KS-95},
according to the symmetry argument of Refs. \cite{cpv-02m,cpv-03r}.
Finally, the $1/n$ expansion of the series  coincides with the expansion close
to four dimensions of the $1/n$ exponents of Refs. \cite{prv-01n,g-02n}.

Since the $\epsilon$ expansion is asymptotic, the series must be properly
resummed to provide results for three-dimensional systems.
The first method that is applied to those series that do not yet behave as 
asymptotic is Pad\'e summation, which consists in extending the series 
analytically by the ratio of two polynomials. 
The approximant $[N/M](\t)=\sum_{i=0}^N a_i \t^i/\sum_{j=0}^M b_j\t^j$ is 
determined by imposing the equivalence between its
Taylor coefficients with the ones of the original series.
We also use Pad\'e-Borel-Leroy (PBL) resummation that works as follows. 
Let $S(\e)=\sum\,s_k\e^k$ be the quantity we want to resum.
The Borel-Leroy transform of $S(\e)$, is defined by 
$B_b(t) = \sum_k s_k t^k/\Gamma(k + b + 1)$.
In order to evaluate $S(\e=1)$ it is necessary to 
perform an analytic continuation of $B_b(t)$, that may be achieved using 
Pad\'e approximants $[N/M]$.
Named $P_S(b,N,M)(t)$ such analytic continuation, estimates of $S(\e)$ are 
given by
\be
\int_0^\infty dt\, t^b e^{-t} P_S(b,N,M)(\e\, t)\,,
\label{intpade}
\ee
depending on the considered Pad\'e $[N/M]$ and on the value of the free 
parameter $b$.
Since the integral (\ref{intpade}) should be defined, the 
approximant must not have poles on the real positive axis.
The Pad\'e with real positive poles are called defective 
and must be discarded in the average procedure.

\section{Analytic expression}
\label{appB}

In this appendix we report some useful analytic expressions of the quantities 
reported in the text. We believe that analytical results 
are very useful for those who want to check our results or to compare 
with other methods where analytic limits are known.

The analytic expressions of Eqs. (\ref{n2+}) are
\begin{eqnarray}
n_0^{\pm}(2) &=& 12 \pm 4\,{\sqrt{6}},\\
n_1^{\pm}(2) &=& -12 \mp {14\over 3}\,{\sqrt{6}},\\
n_2^{\pm}(2)&=& \frac{274 \pm 91\,{\sqrt{6}}}{300}+\frac{( 156 \pm 47\,{\sqrt{6}} ) }{60}\,\zeta(3)\\
n_3^{\pm}(2) &=& \frac{72771 \pm 23369\,{\sqrt{6}}}{67500}+ \frac{( 156 \pm 47\,{\sqrt{6}} ) }{7200}\,{\pi }^4+
  \frac{( 15004 \pm 10621\,{\sqrt{6}} ) }{18000}\,\zeta(3)
\nonumber\\&&
 - \frac{ ( 464 \pm 107\,{\sqrt{6}} )  }{90}\,\zeta(5),\\
n_4^{\pm}(2) &=& \frac{61636782 \pm 14010773\,{\sqrt{6}}}{162000000}+ \frac{( 15004 \pm 10621\,{\sqrt{6}} ) }{2160000}\,{\pi }^4-
  \frac{ ( 464 \pm 107\,{\sqrt{6}} )  } {68040}\,{\pi }^6\nonumber \\
 &&- \frac{86358388 \pm 29909737\,{\sqrt{6}}}{21600000}\,\zeta(3)-  
\frac{7\,\left( 84368 \pm 38339\,{\sqrt{6}} \right) }{720000}\,\zeta(3)^2
\nonumber \\&&+ \frac{( 2753732 \pm 735873\,{\sqrt{6}} ) }
    {432000}\,\zeta(5)
+  \frac{49\,( 1332 \pm 119\,{\sqrt{6}} ) }{6400}\,\zeta(7),
\end{eqnarray}
and for Eqs. (\ref{n3+})
\begin{eqnarray}
n_0^{\pm}(3) &=& 17 \pm 4\,{\sqrt{15}},\\
n_1^{\pm}(3)&=&-17 \mp \frac{259\,{\sqrt{15}}}{60},\\
n_2^{\pm}(3)&=&\frac{2267955 \pm 461534\,{\sqrt{15}}}{2323200}+\frac{42741 \pm 10183\,{\sqrt{15}}}{10560}\,\zeta(3),\\
n_3^{\pm}(3)&=&\frac{141926963025 \pm 31936343543\,{\sqrt{15}}}{101198592000}+\frac{42741 \pm 10183\,{\sqrt{15}}}{1267200}\,\pi^4
\nonumber\\&&+\frac{184976182 \pm 103127153\,{\sqrt{15}}}{306662400}\,\zeta(3) 
-\frac{4835795 \pm 1072837\,{\sqrt{15}}}{696960}\,\zeta(5),\\
n_4^{\pm}(3)&=& \frac{17531276380245840 \pm 3847065393512279\,{\sqrt{15}}}{47020913786880000}-\frac{4835795 \pm 1072837\,{\sqrt{15}}}{526901760}\,\pi^6
\nonumber\\&&
+\frac{184976182 \pm 103127153\,{\sqrt{15}}}{36799488000}\,\pi^4 
-\frac{41214520769938 \pm 9458643900547\,{\sqrt{15}}}{7124380876800}\,\zeta(3)
\nonumber \\&&
+\frac{1871364718325 \pm 375167813937\,{\sqrt{15}}}{161917747200}\,\zeta(5)+\frac{49( 210903545 \pm 35255789\,{\sqrt{15}})}{1090355200}\,\zeta(7)
\nonumber \\&&
+\frac{739213610260 \pm 238830577907\,{\sqrt{15}}}{539725824000}\,\zeta(3)^2\,.
\eea
Similarly for (\ref{etan+}), i.e. the exponents for $m=2$ and  $n=n^+(2,4-\e)$,
\begin{eqnarray}
\eta_{n=n^+} &=&\frac{1}{48}\,\epsilon^2 +\frac{5}{288}\,\epsilon^3
+\left(\frac{-119 + 24\,{\sqrt{6}}}{14400}\zeta(3)+\frac{3691 + 124\,{\sqrt{6}}}{360000} \right) \,\epsilon^4
\nonumber \\&&
+\left( \frac{6408547 + 408608\,{\sqrt{6}}}{1296000000}+
\frac{-119 + 24\,{\sqrt{6}}}{1728000}\,\pi^4
\right.\nonumber \\&&\left.
+\frac{-324561 + 51976\,{\sqrt{6}}}{32400000}\zeta(3)
+\frac{5739 - 1508\,{\sqrt{6}}}{259200} \zeta(5)\right)\e^5+O(\e^6)\,,
\end{eqnarray}
and for (\ref{nun+})
\begin{eqnarray}
{1 \over \nu_{n=n^+}}&=&2-{\e \over 2}+\frac{-1 + {\sqrt{6}}}{50}\e^2+\left(\frac{37 + 4\,{\sqrt{6}}}{1000}\zeta(3)+\frac{397 - 37\,{\sqrt{6}}}{15000}\right)\e^3+ \left(\frac{1104341 - 315351\,{\sqrt{6}}}{18000000}\right.\nonumber \\
&&\left. +\frac{37 + 4\,{\sqrt{6}}}{120000}\pi^4+\frac{-112429 + 46574\,{\sqrt{6}}}{1800000}\zeta(3)-\frac{2473 + 484\,{\sqrt{6}}}{36000}\zeta(5)\right)\e^4  \nonumber \\
&&+\left(\frac{19528133 - 6267483\,{\sqrt{6}}}{270000000}+
 \frac{-112429 + 46574\,{\sqrt{6}}}{216000000}\pi^4-\frac{2473 +484\,{\sqrt{6}}}{27216000}\pi^6 \nonumber \right.\\
&&\left.-\frac{31404323 - 14177332\,{\sqrt{6}}}{216000000}\zeta(3)
+\frac{76622 - 74335\,{\sqrt{6}}}{4500000}\zeta(3)^2 \right. \nonumber \\
&& \left.+\frac{4966979 - 1949164\,{\sqrt{6}}}{21600000}\zeta(5) + \frac{49\,\left( 135 + 74\,{\sqrt{6}} \right) }{80000}\zeta(7) \right)\e^5+O(\e^6) \,.
\end{eqnarray}

\begin{table}[!tbp]
\squeezetable
\caption{\small Four-loop coefficients for the RG functions $\beta_u$ and $\beta_v$.}
\renewcommand\arraystretch{0.3}
\begin{tabular}{c|l}
\multicolumn{1}{c|}{$k,i$}&
\multicolumn{1}{c}{$U_{k,i}(m,n)$}\\
\hline
\hline
5,0&$(17027 - 51250m + 17042m^2 + 154m^3 + 38908m n - 26282 m^2\,n - 
   284\,m^3\,n + 4509\,m^2\,n^2 + 222\,m^3\,n^2 - 46\,m^3\,n^3)/62208$\\
&$+\pi^4(133 - 436\,m + 170\,m^2 + 331\,m\,n - 226\,m^2\,n + 28\,m^2\,n^2)/311040$ \\
&$+\zeta(3)(921 - 2510\,m + 714\,m^2 - 46\,m^3 + 1763\,m\,n - 1096\,m^2\,n + 
   80\,m^3\,n + 214\,m^2\,n^2 - 46\,m^3\,n^2 + 6\,m^3\,n^3)/ 5184$\\
&$+\zeta(5)(1080 - 2820\,m + 660\,m^2 + 1895\,m\,n - 970\,m^2\,n + 155\,m^2\,n^2)/ 5184$ \\
\hline
4,1&$(-340249 + 964828\,m - 283052\,m^2 - 1278\,m^3 - 703072\,m\,n + 
   438218\,m^2\,n + 3098\,m^3\,n - 75927\,m^2\,n^2 $\\
&$- 3312\,m^3\,n^2 + 746\,m^3\,n^3)/248832$\\
&$+\pi^4 (37 + m - 75\,m^2 - 45\,m\,n + 89\,m^2\,n - 7\,m^2\,n^2)/77760 $\\
&$+\zeta(3)(-5511 + 14024 m - 3088\,m^2 + 86\,m^3 - 9401\,m\,n + 4968\,m^2\,n
 -190 m^3 n - 1016 m^2 n^2 + 152 m^3 n^2 - 24 m^3 n^3)/5184$\\
&$+\zeta(5)(-3615 + 8790\,m - 1560\,m^2 - 5575\,m\,n + 2360\,m^2\,n - 400\,m^2\,n^2)/2592$  \\
\hline
3,2&$(92919- 234952 m + 49028 m^2 + 86 m^3 + 159500 m n - 83697 m^2 n - 
   351 m^3 n + 17059 m^2 n^2 + 551 m^3 n^2 - 143 m^3 n^3)/31104$\\
&$+\pi^4 (-143 + 294m - 8 m^2 - 163 m n + 32 m^2 n - 12 m^2 n^2)/38880$\\
&$+\zeta(3)(6404- 15260\,m + 2472\,m^2 - 20\,m^3 + 9784\,m\,n - 4381\,m^2\,n + 
   73\,m^3\,n + 999\,m^2\,n^2 - 89\,m^3\,n^2 + 18\,m^3\,n^3)/2592$\\
&$+\zeta(5)(1225 - 2750\,m + 300\,m^2 + 1625\,m\,n - 500\,m^2\,n + 100\,m^2\,n^2)/324$\\
\hline
2,3&$(-214362 + 477908 m - 49184 m^2 - 296261\,m\,n + 114404\,m^2\,n + 
   210 m^3 n - 32306 m^2\,n^2 - 608 m^3 n^2 + 199\,m^3\,n^3)/62208$\\
&$+ \pi^4(114 - 244\,m + 16\,m^2 + 143\,m\,n - 42\,m^2\,n + 13\,m^2\,n^2)/19440$\\
&$+\zeta(3)(-3854+8452\,m - 744\,m^2 - 5096\,m\,n + 1749\,m^2\,n - 9\,m^3\,n - 
   513\,m^2\,n^2 + 21\,m^3\,n^2 - 6\,m^3\,n^3)/1296$\\
&$+\zeta(5)(-1585 + 3330\,m - 160\,m^2 - 1845\,m\,n + 360\,m^2\,n - 100\,m^2\,n^2)/324$\\
\hline
1,4&$ ( 1 - 2\,m + m\,n) ((31934 + 3957\,m\,n - 14\,m^2\,n^2)/15552 
-\pi^4 (170 + 19\,m\,n)/38880$\\
&\hspace{2.1cm} $+\zeta(3)(4856 + 552\,m\,n + 3\,m^2\,n^2)/2592+
 \zeta(5)(480 + 25\,m\,n)/162)$\\
\hline
0,5&$(-196648 - 80456\,m\,n - 6320\,m^2\,n^2 + 5\,m^3\,n^3)/124416
  +\pi^4(176 + 62\,m\,n + 5\,m^2\,n^2)/38880$\\
&$+\zeta(3)(-2332 - 764\,m\,n - 63\,m^2\,n^2)/1296
 +\zeta(5)(-930 - 275\,m\,n - 10\,m^2\,n^2)/324$\\
\hline
\hline
\multicolumn{1}{c|}{$k,i$}&
\multicolumn{1}{c}{$V_{k,i}(m,n)$}\\
\hline
\hline
5,0&$(-616929+892812 m - 105172 m^2 - 86 m^3 -321700 m n+ 
   63034 m^2 n - 94 m^3 n + 189 m^2 n^2 $\\
&$+ 640 m^3 n^2-174 m^3 n^3)/248832$\\
&$+\pi^4(1409 - 1886\,m + 286\,m^2 + 731\,m\,n - 275\,m^2\,n + 32\,m^2\,n^2)/ 155520$\\
&$+\zeta(3)(-18349 + 20756\,m - 2402\,m^2-12\,m^3- 6203\,m\,n + 1720\,m^2\,n + 
   36\,m^3\,n - 166\,m^2\,n^2 - 36\,m^3\,n^2 + 6\,m^3\,n^3)/5184$\\
&$+\zeta(5)(-32250 + 35760\,m - 1920\,m^2 - 9975\,m\,n + 980\,m^2\,n - 5\,m^2\,n^2)/5184$\\
\hline
4,1&$ (275575 - 341500\,m + 28538\,m^2 + 12\,m^3+ 120765\,m\,n-18548\,m^2\,n + 
   22 m^3 n + 89 m^2\,n^2 - 240 m^3\,n^2 + 87 m^3\,n^3)/31104$\\
&$+\pi^4(-5004 + 5732\,m - 648\,m^2 - 2147\,m\,n + 784\,m^2\,n - 112\,m^2\,n^2)/155520$\\
&$+\zeta(3)(32300 - 32284\,m + 2730\,m^2 + 6\,m^3 + 9321\,m\,n -2380\,m^2\,n - 
   36\,m^3\,n + 312\,m^2\,n^2 + 54\,m^3\,n^2 - 12\,m^3\,n^3)/2592$\\
&$+\zeta(5)(54940 - 55940\,m + 2160\,m^2 + 15225\,m\,n - 1160\,m^2\,n)/2592$\\
\hline
3,2&$(-797194 + 778564\,m - 34544\,m^2 - 278829\,m\,n + 29972\,m^2\,n - 
   22\,m^3\,n - 728\,m^2\,n^2 + 480\,m^3\,n^2 - 261\,m^3\,n^3)/62208$\\
&$+\pi^4(1806 - 1660\,m + 96\,m^2 + 607\,m\,n - 194\,m^2\,n + 38\,m^2\,n^2)/38880$\\
&$+\zeta(3)(-23278+18624\,m-836\,m^2 - 5361\,m\,n + 1149\,m^2\,n + 9\,m^3\,n - 
   222\,m^2\,n^2 - 27\,m^3\,n^2 + 9\,m^3\,n^3)/1296$\\
&$+\zeta(5)(-9545 + 8010\,m - 160\,m^2 - 2190\,m\,n + 120\,m^2\,n)/324$\\
\hline
2,3&$(294788 - 179096\,m + 75554\,m\,n - 4808\,m^2\,n + 269\,m^2\,n^2 - 
   80\,m^3\,n^2 + 87\,m^3\,n^3)/31104 $\\
&$+\pi^4(-340 + 190\,m - 86\,m\,n + 17\,m^2\,n - 6\,m^2\,n^2)/9720$\\
&$+\zeta(3)(17368 - 8416\,m + 2970\,m\,n - 402\,m^2\,n + 144\,m^2\,n^2 + 9\,m^3\,n^2 - 
   6\,m^3\,n^3)/1296$\\
&$+\zeta(5)(1735 - 880\,m + 295\,m\,n - 10\,m^2\,n)/81$\\
\hline
1,4&$ (-157432 - 27832\,m\,n - 56\,m^2\,n^2 - 29\,m^3\,n^3)/41472+
    \pi^4(90 + 17\,m\,n + m^2\,n^2)/6480  $\\
&$+\zeta(3)(-4536 - 560\,m\,n - 24\,m^2\,n^2 + m^3\,n^3)/864
   -\zeta(5) (910 + 105\,m\,n)/108$\\
\end{tabular}
\label{t1}
\end{table}

\begin{table}[!tbp]
\squeezetable
\caption{\small  Five-loop coefficients for the RG function $\beta_u$.}
\renewcommand\arraystretch{0.3}
\begin{tabular}{c|l}
\multicolumn{1}{c|}{$k,i$}&
\multicolumn{1}{c}{$U_{k,i}(m,n)$}\\
\hline
\hline
6,0&$(7720646 - 26391892 m + 11963785 m^2 - 1016454m^3 + 3269m^4 + 
   22679912m\,n - 20725184\,m^2\,n + 1764720\,m^3\,n $\\
&$- 7468\,m^4\,n +4784138\,m^2\,n^2 - 816092\,m^3\,n^2 + 9215\,m^4\,n^2 + 37192\,m^3\,n^3 - 
   6558\,m^4\,n^3 + 771\,m^4\,n^4)/11943936$\\
&$+\pi^4(6643 - 19108\,m + 3786\,m^2 + 2284\,m^3 - 248\,m^4 + 10402\,m\,n + 
   3006\,m^2\,n - 5222\,m^3\,n + 520\,m^4\,n - 5395\,m^2\,n^2 $\\
&$+ 4462\,m^3\,n^2 - 464\,m^4\,n^2 - 882\,m^3\,n^3 + 240\,m^4\,n^3 - 
   24\,m^4\,n^4)/14929920$\\
&$+\pi^6(2705 - 9140 m + 3520\,m^2 + 210\,m^3 + 6950\,m\,n - 4155\,m^2\,n - 
   605 m^3 n + 40 m^2\,n^2 + 555 m^3\,n^2 - 80 m^3\,n^3)/15676416$\\
&$+\zeta(3) (180862 -544185\,m +204185\,m^2-21939\,m^3+215\,m^4+417829\,m\,n - 
   332055\,m^2\,n + 40996\,m^3\,n - 414\,m^4\,n $\\
&$+ 76436\,m^2\,n^2 - 
   25305\,m^3\,n^2 + 303\,m^4\,n^2 + 3187\,m^3\,n^3 - 126\,m^4\,n^3 + 
   11\,m^4\,n^4)/248832$\\
&$+\zeta(3)^2(-37 + 1048 m - 1292 m^2 + 318m^3-1516m\,n + 2571m^2\,n - 
   587m^3\,n - 822m^2\,n^2 + 365\,m^3\,n^2 - 48\,m^3\,n^3)/41472 $\\
&$+\zeta(5)(269531-852440\,m+353290 m^2 -39912 m^3 +701125\,m\,n -627204\,m^2\,n+
   77394\,m^3\,n $\\
&$+ 160298\,m^2\,n^2 - 46682\,m^3\,n^2 + 4600\,m^3\,n^3)/248832 $\\
&$+\zeta(7)(79331 - 228438\,m + 69776\,m^2 + 167139\,m\,n - 105840\,m^2\,n + 
   18032\,m^2\,n^2)/55296$\\
\hline
5,1& $(-23038930 + 73796176\,m - 29373553\,m^2 + 1658632\,m^3 - 3395\,m^4 - 
   60585772\,m\,n + 50569562\,m^2\,n - 3204600\,m^3\,n $\\
&$+ 10406\,m^4\,n - 
   11479472\,m^2\,n^2 + 1780134\,m^3\,n^2 - 17199\,m^4\,n^2 - 
   124124\,m^3\,n^3 + 14082\,m^4\,n^3 - 1947\,m^4\,n^4)/5971968$\\
&$+\pi^4(23753 - 63434\,m + 21402\,m^2 - 5710\,m^3 + 236\,m^4 + 55489\,m\,n - 
   60768\,m^2\,n + 13876\,m^3\,n - 652\,m^4\,n + 25316\,m^2\,n^2 $\\
&$- 
   12066\,m^3\,n^2 + 800\,m^4\,n^2 + 2202\,m^3\,n^3 - 504\,m^4\,n^3 + 
   60\,m^4\,n^4)/7464960$\\
&$+\pi^6(16910 - 23540\,m - 5480\,m^2 - 4800\,m^3 +13575\,m\,n-15650\,m^2\,n + 
   12040\,m^3\,n + 15625\,m^2\,n^2 $\\
&$- 10120\,m^3\,n^2 + 1440\,m^3\,n^3 )/47029248 $\\
&$+\zeta(3)(-1176575 + 3390386\,m - 1109336\,m^2 + 72418\,m^3 - 318\,m^4 - 
   2523565\,m\,n + 1806256\,m^2\,n - 150316\,m^3\,n $\\
&$+ 804\,m^4\,n - 
   402062\,m^2\,n^2 + 108034\,m^3\,n^2 - 830\,m^4\,n^2 - 15286\,m^3\,n^3 + 
   436\,m^4\,n^3 - 46\,m^4\,n^4)/248832$\\
&$+\zeta(3)^2(-28978 + 57820\,m + 4744\,m^2 - 4608\,m^3 - 24477\,m\,n - 17978\,m^2\,n + 
   9112\,m^3\,n $\\
&$+ 9733\,m^2\,n^2 - 6232\,m^3\,n^2 + 864\,m^3\,n^3)/124416$\\
&$+\zeta(5)(-406105 + 1201928\,m - 432714\,m^2 + 42996\,m^3 - 947945\,m\,n + 
   783954\,m^2\,n - 89992\,m^3\,n - 206134\,m^2\,n^2 $\\
&$+ 61028\,m^3\,n^2 - 
   7016\,m^3\,n^3)/62208$\\
&$+\zeta(7)(-44786 + 124852\,m -35280\,m^2 - 89621\,m\,n +54390 m^2 n - 9555\,m^2\,n^2)/4608$\\
\hline
4,2&$(115482976 - 343572912\,m + 116333720\,m^2 - 3730368\,m^3 + 3608\,m^4 + 
   268202738\,m\,n - 201574416\,m^2\,n + 8761141\,m^3\,n $\\
&$- 19289\,m^4\,n + 
   45741288\,m^2\,n^2 - 6288827\,m^3\,n^2 + 46947\,m^4\,n^2 + 
   652565\,m^3\,n^3 - 47076\,m^4\,n^3 + 7905\,m^4\,n^4)/11943936$\\
&$+\pi^4(-267336 + 708928\,m - 194576\,m^2 + 20544\,m^3 - 224\,m^4 - 551881\,m\,n + 
   449912\,m^2\,n - 56150\,m^3\,n + 1072\,m^4\,n $\\
&$- 153429\,m^2\,n^2 + 
   53522\,m^3\,n^2 - 2000\,m^4\,n^2 - 9774\,m^3\,n^3 + 1632\,m^4\,n^3 - 
   240\,m^4\,n^4)/14929920$\\
&$+\pi^6(-174840 + 412360\,m - 71640\,m^2 + 8960\,m^3 - 281095\,m\,n + 
   174430\,m^2\,n - 24600\,m^3\,n - 62815\,m^2\,n^2 $\\
&$+ 22840\,m^3\,n^2 - 
   3600\,m^3\,n^3)/47029248$\\
&$+ \zeta(3)(6054444 - 16659432\,m + 4720608\,m^2 - 170240\,m^3 + 176\,m^4 + 
   12109993\,m\,n - 7988808\,m^2\,n + 429072\,m^3\,n $\\
&$- 818\,m^4\,n + 
   1824513\,m^2\,n^2 - 382176\,m^3\,n^2 + 1350\,m^4\,n^2 + 62156\,m^3\,n^3 - 
   968\,m^4\,n^3 + 130\,m^4\,n^4)/497664$\\
&$+ \zeta(3)^2(  98640 - 215480\,m + 11160\,m^2 + 7040\,m^3 + 111485\,m\,n + 8998\,m^2\,n - 
   16488\,m^3\,n - 16963\,m^2\,n^2 $\\
&$+ 13768\,m^3\,n^2-2160\,m^3\,n^3)/124416$\\
&$+\zeta(5)(1143847 - 3112678\,m + 887044\,m^2 - 62060\,m^3 + 2306576\,m\,n - 
   1653546\,m^2\,n + 153072\,m^3\,n + 446461\,m^2\,n^2 $\\
&$- 126420\,m^3\,n^2 + 
   17704\,m^3\,n^3)/62208$\\
&$+ \zeta(7)( 125489 - 331338\,m + 80360\,m^2+229957\,m\,n - 128576\,m^2\,n + 
   24108\,m^2\,n^2)/4608$\\
\hline
3,3&
$(-41026088 + 109185000\,m - 27504368\,m^2 + 371544\,m^3 - 79647772\,m\,n +
   51460526\,m^2\,n - 1351466\,m^3\,n+ 1484\,m^4\,n $\\
&$- 12678205\,m^2\,n^2 + 
   1407142\,m^3\,n^2 - 6890\,m^4\,n^2 - 218338\,m^3\,n^3 + 9456\,m^4\,n^3 - 
   2025\,m^4\,n^4)/2985984$\\
&$ +\pi^4(62914 - 161580\,m + 37328\,m^2-1576\,m^3+120367\,m\,n-85326\,m^2\,n 
+ 6208\,m^3\,n - 36\,m^4\,n + 27731\,m^2\,n^2$\\
&$ - 7596\,m^3\,n^2 +  132\,m^4\,n^2 + 1560\,m^3\,n^3 - 156\,m^4\,n^3 + 30\,m^4\,n^4)/1866240 $\\
&$+\pi^6(23335 - 55490 m + 9140 m^2 - 320\,m^3 + 37255\,m\,n -20300\,m^2\,n+ 
   1280\,m^3\,n + 6360\,m^2\,n^2 - 1560\,m^3\,n^2 $\\
&$+ 300\,m^3\,n^3)/2939328 $\\
&$+\zeta(3)(-1055134 + 2705500\,m - 604488\,m^2 + 9256\,m^3 - 1901069\,m\,n + 
   1132320\,m^2\,n - 35686\,m^3\,n + 4\,m^4\,n $\\
&$- 285779\,m^2\,n^2 + 
   43744\,m^3\,n^2 - 18\,m^4\,n^2 - 8669\,m^3\,n^3 + 24\,m^4\,n^3 - 
   5\,m^4\,n^4)/62208 $\\
&$-\zeta(3)^2(10409 - 22174 m + 1132 m^2  +224 m^3+11441 m n+92m^2 n - 
   800 m^3 n - 1080m^2 n^2 + 936 m^3 n^2 - 180 m^3 n^3)/7776 $\\
&$+\zeta(5)(-454646 + 1130856\,m - 229244\,m^2 + 7680\,m^3 - 781415\,m\,n + 
   460582\,m^2\,n - 28608\,m^3\,n - 132069\,m^2\,n^2 $\\
&$+ 32800\,m^3\,n^2 - 
   5936\,m^3\,n^3)/15552$\\
&$ +\zeta(7)(-36946 + 89572\,m - 15680\,m^2 - 58653\,m\,n + 27734\,m^2\,n - 6027\,m^2\,n^2)/864$ \\
\hline
2,4&$(140915584 - 325697728\,m + 43866560\,m^2 + 217044656\,m\,n - 
   109698320\,m^2\,n + 1306736\,m^3\,n + 34192982\,m^2\,n^2 $\\
&$- 
   2560000\,m^3\,n^2 + 5796\,m^4\,n^2 + 633775\,m^3\,n^3 - 14286\,m^4\,n^3 + 
   4245\,m^4\,n^4)/11943936$\\
&$+\pi^4(-65666 + 153224\,m - 21892\,m^2 - 106391\,m\,n + 61604\,m^2\,n - 
   2046\,m^3\,n - 22040\,m^2\,n^2 + 4392\,m^3\,n^2 - 24\,m^4\,n^2 $\\
&$- 
   1200\,m^3\,n^3 + 54\,m^4\,n^3 - 15\,m^4\,n^4)/1866240$\\
&$+\pi^6(-8245+18410\,m-1920\,m^2 - 11770\,m\,n + 5260\,m^2\,n - 130\,m^3\,n - 
   1810\,m^2\,n^2 + 280\,m^3\,n^2 - 75\,m^3\,n^3)/979776$\\
&$+\zeta(3)(3628416-8316352\,m+1059520\,m^2+ 5471024\,m\,n - 2663456\,m^2\,n + 
   37760\,m^3\,n + 843534\,m^2\,n^2 - 83056\,m^3\,n^2 $\\
&$- 76\,m^4\,n^2 + 
   22575\,m^3\,n^3 + 146\,m^4\,n^3 - 35\,m^4\,n^4)/248832$\\
&$+\zeta(3)^2(3409 - 6898\,m + 80\,m^2+3370\,m\,n + 236\,m^2\,n - 78\,m^3\,n - 
   242\,m^2\,n^2 + 168\,m^3\,n^2 - 45\,m^3\,n^3)/2592$\\
&$+\zeta(5)(283674 - 637364\,m + 70016\,m^2 + 407131\,m\,n - 182346\,m^2\,n + 
   5448\,m^3\,n + 61874\,m^2\,n^2 - 11418\,m^3\,n^2 $\\
&$+ 2985\,m^3\,n^3)/10368$\\
&$+\zeta(7)(45815 - 100450\,m + 8820\,m^2 + 60760\,m\,n - 21070\,m^2\,n + 6125\,m^2\,n^2)/1152$\\
\hline
1,5&$ ( 1 - 2 m + m n) (
(-34395968 - 6592896\,m\,n - 119350\,m^2\,n^2 - 483\,m^3\,n^3)/ 5971968$\\
&$ +\pi^4(40708 + 9356\,m\,n + 513\,m^2\,n^2 + 3\,m^3\,n^3)/1866240
+\pi^6(4810 + 815\,m\,n + 30\,m^2\,n^2)/979776$\\
&$
+\zeta(3)(-934072 - 172480\,m\,n - 4080\,m^2\,n^2 + 17\,m^3\,n^3)/124416
+\zeta(3)^2(-1574 + 95\,m\,n + 18\,m^2\,n^2)/2592$\\
&$
+\zeta(5)(-37618 - 6069\,m\,n - 301\,m^2\,n^2)/2592
+\zeta(7)( -5782 - 637\,m\,n)/288)$\\
\hline
0,6&$(13177344 + 6646336\,m\,n + 808496\,m^2\,n^2 + 12578\,m^3\,n^3 + 13\,m^4\,n^4)/3981312$\\
&$+\pi^4(-21120 - 9532\,m\,n - 1388\,m^2\,n^2 - 63\,m^3\,n^3)/1244160
+\pi^6(-7440 - 3130\,m\,n - 355\,m^2\,n^2 - 10\,m^3\,n^3)/1959552$\\
&$+\zeta(3)(1314336 + 552280\,m\,n + 67640\,m^2\,n^2 + 1248\,m^3\,n^3 - 9\,m^4\,n^4)/248832$\\
&$+\zeta(3)^2(3264 + 446\,m\,n - 59\,m^2\,n^2 - 6\,m^3\,n^3)/ 5184
+\zeta(5)(165084 + 66986\,m\,n + 7466\,m^2\,n^2 + 305\,m^3\,n^3)/15552$\\
&$
+\zeta(7)(25774 + 9261\,m\,n + 686\,m^2\,n^2)/1728$\\
\end{tabular}
\label{6loopbu}
\end{table}

\begin{table}[!tbp]
\squeezetable
\caption{\small Five-loop coefficients for the RG function $\beta_v$.}
\renewcommand\arraystretch{0.3}
\begin{tabular}{c|l}
\multicolumn{1}{c|}{$k,i$}&
\multicolumn{1}{c}{$V_{k,i}(m,n)$}\\
\hline
\hline
6,0&$(-72238776 + 132078652m - 26680036m^2 + 273681m^3 - 345\,m^4 - 
   60428904\,m\,n + 22666801m^2\,n + 204824m^3\,n $\\
&$- 533m^4\,n - 
   1463603\,m^2\,n^2 - 234959\,m^3\,n^2 + 4376\,m^4\,n^2 - 3812\,m^3\,n^3 - 
   4626\,m^4\,n^3 + 732\,m^4\,n^4)/ 11943936$\\
&$+\pi^4(283706-451827\,m + 105356\,m^2 - 5907\,m^3 - 24\,m^4 + 200762\,m\,n - 
   107414\,m^2\,n + 6718\,m^3\,n + 96\,m^4\,n $\\
&$+ 17168\,m^2\,n^2 - 
   2650\,m^3\,n^2 - 144\,m^4\,n^2 + 84\,m^3\,n^3 + 96\,m^4\,n^3 
   - 12\,m^4\,n^4)/7464960$\\
&$+\pi^6(435460 - 674770\,m + 130820\,m^2 - 5040\,m^3 + 288075\,m\,n - 
  128455\,m^2\,n + 4790\,m^3\,n $\\
&$+ 16985\,m^2\,n^2 - 1175\,m^3\,n^2)/47029248$\\
&$+\zeta(3)(-5923864 + 8822466\,m - 1755312\,m^2 + 19168\,m^3 + 334\,m^4 - 
   3429421\,m\,n + 1453852\,m^2\,n - 4514\,m^3\,n $\\
&$- 734\,m^4\,n - 
   170929\,m^2\,n^2 - 6198\,m^3\,n^2 + 752\,m^4\,n^2 + 2384\,m^3\,n^3 - 
   388\,m^4\,n^3 + 40\,m^4\,n^4)/ 497664$\\
&$+\zeta(3)^2(-317108+284654m-6772m^2-2640m^3-40197 m n- 14179m^2\,n + 
   2426m^3\,n + 3431m^2 n^2 - 641m^3\,n^2)/124416$\\
&$+\zeta(5)(-6417032 +9071536\,m - 1459068\,m^2 + 81384\,m^3 - 3393769\,m\,n + 
   1212152\,m^2\,n - 67088\,m^3\,n $\\
&$- 133666\,m^2\,n^2 + 16070\,m^3\,n^2 - 
   608\,m^3\,n^3)/248832$\\
&$+\zeta(7)(-2229108 + 2985472\,m - 375144\,m^2 - 1019788\,m\,n + 264208\,m^2\,n - 
   18277\,m^2\,n^2)/55296$\\
\hline
5,1&$(77055030 - 125871804 m + 21294903 m^2 - 149415 m^3 + 96 m^4 + 
   55861516 m\,n - 19092239\,m^2\,n - 180275\,m^3\,n + 262\,m^4\,n $\\
&$+ 
   1334594\,m^2\,n^2 + 213766\,m^3\,n^2 - 3282\,m^4\,n^2 - 4687\,m^3\,n^3 + 
   4626\,m^4\,n^3 - 915\,m^4\,n^4)/2985984$\\
&$+\pi^4(-620803 + 882404\,m -171586\,m^2+6676\,m^3 + 12\,m^4 - 381314\,m\,n + 
   192706\,m^2\,n - 10104\,m^3\,n - 96\,m^4\,n $\\
&$- 33043\,m^2\,n^2 + 
   5296\,m^3\,n^2 + 216\,m^4\,n^2 - 324\,m^3\,n^3 - 192\,m^4\,n^3 + 
   30\,m^4\,n^4)/3732480$\\
&$+\pi^6(-936130 + 1316400\,m - 210360\,m^2 + 5600\,m^3 - 548055\,m\,n + 
   230560\,m^2\,n - 6720\,m^3\,n $\\
&$- 33685\,m^2\,n^2 +2120\,m^3\,n^2)/23514624$\\
&$+\zeta(3)(6309091 - 8414600\,m + 1418328\,m^2 - 10638\,m^3 - 88\,m^4 + 
   3170292\,m\,n - 1282310\,m^2\,n + 2934\,m^3\,n $\\
&$+ 364\,m^4\,n + 
   168641\,m^2\,n^2 + 5428\,m^3\,n^2 - 564\,m^4\,n^2 - 2348\,m^3\,n^3 + 
   388\,m^4\,n^3 - 50\,m^4\,n^4)/124416$\\
&$+\zeta(3)^2(619622 - 535776m + 13176m^2 + 2912m^3 + 65193m n + 30760 m^2 n - 
   3360m^3\,n - 7237m^2\,n^2 + 1208 m^3 n^2)/ 62208$\\
&$+\zeta(5)(3418732 - 4467538\,m + 584784\,m^2 - 22800\,m^3 + 1633807\,m\,n - 
   535301\,m^2\,n + 22976\,m^3\,n $\\
&$+ 66871\,m^2\,n^2 - 7352\,m^3\,n^2 + 
   324\,m^3\,n^3)/ 31104$\\
&$+\zeta(7)(766066 - 958440\,m + 101528\,m^2 + 318647\,m\,n - 75068\,m^2\,n + 
   5733\,m^2\,n^2)/4608$\\
\hline
4,2&$(-139217840 +194720568\,m - 24121596\,m^2 + 91684\,m^3 - 83985744\,m\,n + 
   24543584\,m^2\,n + 200423\,m^3\,n - 131\,m^4\,n $\\
&$- 2334581\,m^2\,n^2 - 
   291418\,m^3\,n^2 + 3282\,m^4\,n^2 + 23336\,m^3\,n^3 - 6939\,m^4\,n^3 + 
   1830\,m^4\,n^4)/2985984$\\
&$+\pi^4( 2301104 - 2794568\,m + 402488\,m^2 - 7856\,m^3 + 1178888\,m\,n - 
   545040\,m^2\,n + 20644\,m^3\,n + 96\,m^4\,n + 104065\,m^2\,n^2 $\\
&$- 
   15916\,m^3\,n^2 - 432\,m^4\,n^2 + 1632\,m^3\,n^3 + 576\,m^4\,n^3 - 
   120\,m^4\,n^4)/7464960$\\
&$+\pi^6(3382640 - 4147520\,m + 486480\,m^2 - 6400\,m^3 + 1684280\,m\,n - 
   650500\,m^2\,n + 13040\,m^3\,n $\\
&$+ 109785m^2\,n^2-5960\,m^3\,n^2)/47029248$\\
&$+\zeta(3)(-22967160 + 26065200\,m - 3253224\,m^2 +13408\,m^3-9525746\,m\,n + 
   3532984\,m^2\,n - 5288\,m^3\,n - 364\,m^4\,n $\\
&$- 553459\,m^2\,n^2 - 
   15716\,m^3\,n^2 + 1128\,m^4\,n^2 + 7540\,m^3\,n^3 - 1164\,m^4\,n^3 + 
   200\,m^4\,n^4)/248832$\\
&$+\zeta(3)^2(-2079760 + 1637632\,m - 28080\,m^2 - 3328\,m^3 - 169288\,m\,n - 
   106228\,m^2\,n + 6416\,m^3\,n $\\
&$+ 24069\,m^2\,n^2 - 3512\,m^3\,n^2)/124416$\\
&$+\zeta(5)(-12244940+14181036\,m -1355856\,m^2 + 26480\,m^3 - 5088273\,m\,n + 
   1480496\,m^2\,n - 43840\,m^3\,n $\\
&$- 219687\,m^2\,n^2 + 21140\,m^3\,n^2 - 
   1128\,m^3\,n^3)/62208$\\
&$+\zeta(7)(-1342404 + 1494108\,m-115640\,m^2 - 484267\,m\,n + 98196\,m^2\,n - 
   9408\,m^2\,n^2)/4608$\\
\hline
3,3&
$(67986200 - 75382408\,m + 5048536\,m^2 + 32379740\,m\,n - 7295478\,m^2\,n - 34186\,m^3\,n + 1144333\,m^2\,n^2 + 88907\,m^3\,n^2 $\\
&$ - 547\,m^4\,n^2 -  16178\,m^3\,n^3 + 2313\,m^4\,n^3 - 915\,m^4\,n^4)/1492992$\\
&$+\pi^4(-581572 + 554120\,m - 43136\,m^2 - 236664\,m\,n + 92168\,m^2\,n - 
   1808\,m^3\,n - 21599\,m^2\,n^2 + 2670\,m^3\,n^2 + 36\,m^4\,n^2  $\\
&$-  462\,m^3\,n^3 - 96\,m^4\,n^3 + 30\,m^4\,n^4)/1866240 $\\
&$+\pi^6(-207670 + 203340\,m - 12800\,m^2 - 83285\,m\,n + 27270\,m^2\,n - 
   280\,m^3\,n - 5880\,m^2\,n^2 + 240\,m^3\,n^2)/2939328 $\\ 
&$+\zeta(3)(5715252 - 5053816\,m + 339728\,m^2 +1855772\,m\,n-572300\,m^2\,n + 
   428\,m^3\,n + 119493\,m^2\,n^2 + 2696\,m^3\,n^2  $\\
&$- 94\,m^4\,n^2 - 1591\,m^3\,n^3 + 194\,m^4\,n^3 - 50\,m^4\,n^4)/ 62208$\\
&$+\zeta(3)^2(120242-76548\,m+448\,m^2+6379\,m\,n+5334\,m^2\,n - 136\,m^3\,n - 
   1296\,m^2\,n^2 + 144\,m^3\,n^2)/ 7776 $\\
&$+\zeta(5)(2982544 - 2773064\,m + 145016\,m^2 +1014902\,m\,n-244024\,m^2\,n + 
   3780\,m^3\,n + 47575\,m^2\,n^2 - 3482\,m^3\,n^2 $\\
&$+ 254\,m^3\,n^3)/15552 $\\
&$+\zeta(7)(240933 - 216090\,m + 8820\,m^2 + 70560\,m\,n - 11270\,m^2\,n + 
   1568\,m^2\,n^2)/864$\\
\hline
2,4&$ (-152904896 + 107048320\,m - 53120672\,m\,n + 7504128\,m^2\,n - 
   2268736\,m^2\,n^2 - 82186\,m^3\,n^2 + 36434\,m^3\,n^3 $\\
&$- 2313\,m^4\,n^3 + 
   1830\,m^4\,n^4)/ 5971968$\\
&$+\pi^4(340984 - 196700m + 106176m n - 26092 m^2 n + 9902 m^2\,n^2 - 
   675 m^3 n^2 + 243\,m^3 n^3 + 24 m^4 n^3 - 15m^4 n^4)/ 1866240$\\
&$+\pi^6(39720 - 23650\,m + 12170\,m\,n - 2535\,m^2\,n + 910\,m^2\,n^2 - 20\,m^3\,n^2)/979776$\\
&$+\zeta(3)(-6552944 + 3545400\,m - 1587312\,m\,n + 309792\,m^2\,n 
   - 110060\,m^2\,n^2 - 1472\,m^3\,n^2 + 1466\,m^3\,n^3 $\\
&$- 97\,m^4\,n^3 + 50\,m^4\,n^4)/124416$\\
&$+\zeta(3)^2( -21632 + 8110\,m - 414\,m\,n - 559\,m^2\,n + 206\,m^2\,n^2 - 12\,m^3\,n^2)/2592$\\
&$+\zeta(5)(-282530 + 159690\,m-73301\,m\,n + 11355\,m^2\,n - 3739\,m^2\,n^2 + 
   148\,m^3\,n^2 - 20\,m^3\,n^3)/ 2592$\\
&$+\zeta(7)(-44835 + 24500\,m- 9800\,m\,n + 980\,m^2\,n - 245\,m^2\,n^2)/288$\\
\hline
1,5&$(24902080 + 6117632\,m\,n + 222434\,m^2\,n^2 - 3557\,m^3\,n^3 - 183\,m^4\,n^4)/2985984$\\
&$+\pi^4(-108664 - 26368\,m\,n - 2110\,m^2\,n^2 - 48\,m^3\,n^3 + 3\,m^4\,n^4)/1866240
+\pi^6 (-6350 - 1475\,m\,n - 95\,m^2\,n^2)/489888$\\
&$+\zeta(3)(1037432 + 188904\,m\,n + 11140\,m^2\,n^2 - 151\,m^3\,n^3 - 5\,m^4\,n^4)/62208
+\zeta(3)^2(3322 - 23\,m\,n - 23\,m^2\,n^2)/1296$\\
&$+\zeta(5)(89848 + 17230\,m\,n + 795\,m^2\,n^2 + 4\,m^3\,n^3)/2592
 +\zeta(7)( 14210 + 2205\,m\,n + 49\,m^2\,n^2)/288$\\
\end{tabular}
\label{6loopbv}
\end{table}

\begin{table}[!tbp]
\squeezetable
\caption{\small Four and Five-loop coefficients for the RG function $\eta$.}
\renewcommand\arraystretch{0.3}
\begin{tabular}{c|l}
\multicolumn{1}{c|}{$k,i$}&
\multicolumn{1}{c}{$E_{k,i}(m,n)$}\\
\hline
\hline
4,0&$ -5 ( -397 + 1100 m - 292 m^2 - 14 m^3 - 760 m n + 394 m^2 n +  26 m^3 n - 43 m^2 n^2 - 16 m^3 n^2 + 2 m^3 n^3 )/ 165888$\\
3,1&$ 5 ( -147 + 376 m - 80 m^2 - 2 m^3 - 253 m n + 124 m^2 n +    6 m^3 n - 19 m^2 n^2 - 6 m^3 n^2 + m^3 n^3 )/20736$\\
2,2&$ -5 ( -118 + 268 m - 32 m^2 - 171 m n + 72 m^2 n + 2 m^3 n -    18 m^2 n^2 - 4 m^3 n^2 + m^3 n^3 )/ 13824$\\
1,3&$ 5 ( 1 - 2 m + m n )   ( -100 - 18 m n + m^2 n^2 )/ 20736$\\
0,4&$ -5 ( 2 + m n )  ( -100 - 18 m n + m^2 n^2 )/82944$\\
\hline
5,0&$(174432 - 550224\,m + 207810\,m^2 - 6801\,m^3 + 351\,m^4 + 436969\,m\,n - 
   332893\,m^2\,n + 9998\,m^3\,n $\\
&$- 819\,m^4\,n + 63604\,m^2\,n^2 - 2905\,m^3\,n^2 + 780\,m^4\,n^2 + 49\,m^3\,n^3 - 390\,m^4\,n^3 + 
   39\,m^4\,n^4)/5971968$\\
&$+\pi^4(314 - 908\,m + 280\,m^2 + 659\,m\,n - 410\,m^2\,n + 65\,m^2\,n^2)/2488320$\\
&$+\zeta(3)(-1490+5528\,m - 2320\,m^2 - 192\,m^3 - 36\,m^4 -4893\,m\,n+3738\,m^2\,n + 
   436\,m^3\,n + 84\,m^4\,n - 511\,m^2\,n^2 $\\
&$- 316\,m^3\,n^2 - 80\,m^4\,n^2 +  16\,m^3\,n^3 + 40\,m^4\,n^3 - 4\,m^4\,n^4)/ 497664$\\
\hline
4,1&$( -635344 + 1905920 m - 649324 m^2 + 14482 m^3 - 390 m^4 -  1482454 m n + 1083994 m^2 n - 26566 m^3 n + 1560 m^4 n $\\
&$-  222127 m^2 n^2 + 11924 m^3 n^2 - 2340 m^4 n^2 - 700 m^3 n^3 +  1560 m^4 n^3 - 195 m^4 n^4)/5971968$\\
&$+\pi^4(-163 + 460\,m - 134\,m^2 - 331\,m\,n + 202\,m^2\,n - 34\,m^2\,n^2)/373248$\\
&$+\zeta(3)(1289 - 4590\,m + 1894\,m^2 + 108\,m^3 + 10\,m^4 + 4044\,m\,n - 
   3154\,m^2\,n - 304\,m^3\,n - 40\,m^4\,n + 457\,m^2\,n^2 $\\
&$+ 286\,m^3\,n^2 + 60\,m^4\,n^2 - 25\,m^3\,n^3 - 40\,m^4\,n^3 + 5\,m^4\,n^4)/ 124416$\\
\hline
3,2&$( 493416 - 1352120 m + 369376 m^2 - 4088 m^3 + 1014866 m n -  689396 m^2 n + 12174 m^3 n - 390 m^4 n + 163973 m^2 n^2 $\\
&$-  9096 m^3 n^2 + 1170 m^4 n^2 + 1090 m^3 n^3 - 1170 m^4 n^3 +  195 m^4 n^4)/ 2985984$\\
&$+\pi^4(61 - 160\,m + 38\,m^2 + 112\,m\,n- 64\,m^2\,n + 13\,m^2\,n^2)/93312$\\
&$+\zeta(3)(-2032 + 6300\,m - 2172\,m^2 - 64\,m^3 -5357\,m\,n + 4102\,m^2\,n + 
   292\,m^3\,n + 20\,m^4\,n - 736\,m^2\,n^2 - 398\,m^3\,n^2 $\\
&$- 60\,m^4\,n^2 +  55\,m^3\,n^3 + 60\,m^4\,n^3 - 10\,m^4\,n^4)/ 124416$\\
\hline
2,3&$( -417648 + 995568 m - 160272 m^2 - 704492 m n + 417416 m^2 n -  4000 m^3 n - 131402 m^2 n^2 + 6050 m^3 n^2 - 390 m^4 n^2$\\
&$ -  1415 m^3 n^3 + 780 m^4 n^3 - 195 m^4 n^4)/2985984$\\
&$+\pi^4(-50 + 116\,m - 16\,m^2 - 77\,m\,n + 38\,m^2\,n -11\,m^2\,n^2)/93312$\\
&$+\zeta(3)(1856 - 4616\,m + 904\,m^2 +3594\,m\,n-2472\,m^2\,n - 100\,m^3\,n + 
   649\,m^2\,n^2 + 250\,m^3\,n^2 + 20\,m^4\,n^2 $\\
&$- 55\,m^3\,n^3 - 40\,m^4\,n^3 + 10\,m^4\,n^4)/124416$\\
\hline
1,4&$(1 - 2 m + m n)(5( 77056 + 22752 m n + 296m^2n^2 + 39m^3 n^3 )  
+ 64 \pi^4 (22 + 5 m n) $\\
&\hspace{2cm}$-240\zeta(3)(184+64m n - 6m^2n^2 + m^3n^3) )/5971968$\\
\hline
0,5&$-( 2 + m n)(5( 77056 + 22752 m n + 296m^2n^2 + 39m^3 n^3 )
+ 64 \pi^4 (22 + 5 m n)$\\
&\hspace{1.3cm}$- 240\zeta(3)(184+64m n - 6m^2n^2 + m^3n^3))/29859840$\\
\end{tabular}
\label{tabeta}
\end{table}

\begin{table}[tbp]
\squeezetable
\caption{\small Four and five-loop coefficients for the RG function $1/\nu$.}
\renewcommand\arraystretch{0.3}
\begin{tabular}{c|l}
\multicolumn{1}{c|}{$k,i$}&
\multicolumn{1}{c}{$B_{k,i}(m,n)$}\\
\hline
\hline
4,0&$(112697-323980 m + 97772m^2 + 814m^3 + 237400m n - 
   150394m^2n - 426m^3n + 26503m^2n^2 - 384m^3n^2 $\\
&$-2m^3n^3)/497664+\pi^4(163 - 460\,m + 134\,m^2 + 331\,m\,n - 202\,m^2\,n + 34\,m^2\,n^2)/311040$\\
&$+\zeta(3)(273-632\,m+136\,m^2-50\,m^3+373\,m\,n - 196\,m^2\,n + 82\,m^3\,n + 
   52\,m^2\,n^2 - 44\,m^3\,n^2 + 6\,m^3\,n^3)/10368$\\
3,1&$(-42647 + 112736m - 27340m^2 - 102m^3 - 80233m\,n+47674\,m^2\,n + 
   56\,m^3\,n - 10189\,m^2\,n^2 + 44\,m^3\,n^2 + m^3\,n^3)/62208$\\
&$+\pi^4(-61 + 160\,m - 38\,m^2 - 112\,m\,n + 64\,m^2\,n - 13\,m^2\,n^2)/38880$\\
&$+\zeta(3)(-214 + 492\,m - 78\,m^2 + 14\,m^3 - 284\,m\,n + 113\,m^2\,n - 37\,m^3\,n - 
   35\,m^2\,n^2 + 35\,m^3\,n^2 - 6\,m^3\,n^3)/ 2592$\\
2,2&$(35118 - 81708\,m + 11472\,m^2 + 55091\,m\,n- 28472\,m^2\,n - 2\,m^3\,n + 
   8498\,m^2\,n^2 + 4\,m^3\,n^2 - m^3\,n^3)/ 41472$\\
&$+\pi^4(50 - 116\,m + 16\,m^2 + 77\,m\,n - 38\,m^2\,n + 11\,m^2\,n^2)/25920$\\
&$+\zeta(3)(82 - 184\,m + 20\,m^2 + 106\,m\,n - 34\,m^2\,n + 6\,m^3\,n + 13\,m^2\,n^2 - 
   12\,m^3\,n^2 + 3\,m^3\,n^3)/864$\\
1,3&$-( 1 - 2 m + m n )(155300 + 37890 m n - 5 m^2 n^2 
  +\pi^4(352+80 m n) +\zeta(3) (16320 +2400 m n + 720 m^2 n^2 ))/311040$\\
0,4&$( 2 + m n )  (155300 + 37890 m n - 5 m^2 n^2 
  +\pi^4(352+80 m n) +\zeta(3) (16320 +2400 m n + 720 m^2 n^2 ))/1244160$\\
\hline
5,0&$(6716354 - 21727736\,m + 8800405\,m^2 - 507480\,m^3 + 2103\,m^4 + 
   17902846m\,n - 15072442m^2\,n + 995884m^3\,n - 1398\,m^4\,n $\\
&$+3424054\,m^2\,n^2 - 575534\,m^3\,n^2 - 537\,m^4\,n^2 + 43670\,m^3\,n^3 - 
   210\,m^4\,n^3 + 21\,m^4\,n^4)/11943936$\\
&$+\pi^4(24815 - 76670\,m + 28406\,m^2 - 1226\,m^3 - 140\,m^4 + 60313\,m\,n - 
   46884\,m^2\,n + 2660\,m^3\,n + 268\,m^4\,n + 10142\,m^2\,n^2 $\\
&$-    1582\,m^3\,n^2 - 224\,m^4\,n^2 + 14\,m^3\,n^3 + 120\,m^4\,n^3 - 
   12\,m^4\,n^4)/ 14929920$\\
&$+\pi^6(34410 - 105900\,m + 39000\,m^2 -1920\,m^3+82855\,m\,n-63370\,m^2\,n + 
   3560\,m^3\,n + 13165\,m^2\,n^2 $\\
&$- 1960\,m^3\,n^2 + 160\,m^3\,n^3)/94058496$\\
&$+\zeta(3)(156293-443202\,m+147804\,m^2-17482\,m^3+ 294\,m^4 + 320673\,m\,n - 
   231192\,m^2\,n + 33756\,m^3\,n - 708\,m^4\,n $\\
&$+ 52756\,m^2\,n^2 - 
   22810\,m^3\,n^2 + 686\,m^4\,n^2 + 3438\,m^3\,n^3 - 340\,m^4\,n^3 + 
   34\,m^4\,n^4)/ 497664$\\
&$+\zeta^2(3)(-23838+70212\,m-22920\,m^2+384\,m^3-52157\,m\,n+ 34814\,m^2\,n - 
   712\,m^3\,n $\\
&$- 6143\,m^2\,n^2 + 392\,m^3\,n^2 - 32\,m^3\,n^3)/ 248832$\\
&$+\zeta(5)(96 - 671\,m+556\,m^2-77\,m^3+698\,m\,n- 701\,m^2\,n - 24\,m^3\,n - 
   13\,m^2\,n^2 + 171\,m^3\,n^2 - 35\,m^3\,n^3) / 31104$\\
\hline
4,1&$(-24959332 + 76197848\,m - 27373258\,m^2 + 1096198\,m^3 - 2124\,m^4 - 
   61187662\,m\,n + 48929662\,m^2\,n - 2753983\,m^3\,n $\\
&$+ 1797\,m^4\,n - 
   11819002\,m^2\,n^2 + 2081903\,m^3\,n^2 - 303\,m^4\,n^2 - 
   212479\,m^3\,n^3 + 840\,m^4\,n^3 - 105\,m^4\,n^4) / 11943936$\\
&$+\pi^4(-89558+264608\,m - 88316\,m^2 + 2672\,m^3 + 152\,m^4 - 204363\,m\,n + 
   152024\,m^2\,n - 7410\,m^3\,n - 496\,m^4\,n $\\
&$- 35089\,m^2\,n^2 + 
   5766\,m^3\,n^2 + 704\,m^4\,n^2 - 274\,m^3\,n^3 - 480\,m^4\,n^3 + 
   60\,m^4\,n^4) / 14929920$\\
&$+\pi^6(-31085 + 91700\,m -30610\,m^2+1080\,m^3-70435\,m\,n + 51610\,m^2\,n - 
   2440\,m^3\,n $\\
&$- 11380\,m^2\,n^2 + 1760\,m^3\,n^2 -200\,m^3\,n^3)/ 23514624$\\
&$+\zeta(3)(-571026 + 1564720\,m - 462600\,m^2 + 40260\,m^3 - 328\,m^4 - 
   1112181\,m\,n + 750388\,m^2\,n - 92100\,m^3\,n + 1354\,m^4\,n $\\
&$-    182311\,m^2\,n^2 + 78880\,m^3\,n^2 - 2046\,m^4\,n^2 - 14200\,m^3\,n^3 + 
   1360\,m^4\,n^3 - 170\,m^4\,n^4) / 497664$\\
&$+\zeta^2(3)(20887-59740\,m+18182\,m^2-216\,m^3+43877\,m\,n - 28502\,m^2\,n + 
   488\,m^3\,n + 5336\,m^2\,n^2 $\\
&$- 352\,m^3\,n^2 + 40\,m^3\,n^3)/ 62208$\\
&$+\zeta(5)(-736 + 2826\,m - 1524m^2 + 170m^3 - 2516m\,n +2084\,m^2\,n + 
   122\,m^3\,n - 17\,m^2\,n^2 - 526\,m^3\,n^2 + 117\,m^3\,n^3) / 31104$\\
\hline
3,2&$(19701864 - 54683960\,m + 15600904\,m^2 - 320672\,m^3 + 42271184\,m\,n - 
   31187114\,m^2\,n + 1328916\,m^3\,n - 210\,m^4\,n $\\
&$+ 8513717\,m^2\,n^2 - 
   1441854\,m^3\,n^2 + 630\,m^4\,n^2 + 217120\,m^3\,n^3 - 630\,m^4\,n^3 + 
   105\,m^4\,n^4) / 5971968$\\
&$+\pi^4(17255-46902\,m+ 12576\,m^2 - 184\,m^3 + 35060\,m\,n - 24131\,m^2\,n + 
   883\,m^3\,n + 30\,m^4\,n + 6323\,m^2\,n^2 $\\
&$- 1001\,m^3\,n^2 - 
   90\,m^4\,n^2 + 106\,m^3\,n^3 + 90\,m^4\,n^3 - 15\,m^4\,n^4)/ 1866240$\\
&$+\pi^6(11995 - 32600\,m + 8770\,m^2 - 160\,m^3 + 24220\,m\,n-16420\,m^2\,n + 
   580\,m^3\,n $\\
&$+ 4135\,m^2\,n^2 - 620\,m^3\,n^2 + 100\,m^3\,n^3)/ 5878656
$\\
&$+\zeta(3)(218618 - 564220\,m + 133328\,m^2 - 6344\,m^3 + 390068\,m\,n - 
   238048\,m^2\,n + 22302\,m^3\,n - 170\,m^4\,n + 65749\,m^2\,n^2 $\\
&$- 
   27288\,m^3\,n^2 + 510\,m^4\,n^2 + 5920\,m^3\,n^3 - 510\,m^4\,n^3 + 
   85\,m^4\,n^4)/ 124416$\\
&$+\zeta^2(3)(-7889 + 20920 m - 5174 m^2 +32 m^3-14924 m n+9044 m^2 n - 
   116 m^3 n - 1997 m^2 n^2 + 124 m^3 n^2 - 20 m^3 n^3) / 15552$\\
&$+\zeta(5)(1364 - 4444\,m + 1796\,m^2 - 80\,m^3 + 3569\,m\,n - 2486\,m^2\,n - 
   208\,m^3\,n + 38\,m^2\,n^2 + 614\,m^3\,n^2 - 163\,m^3\,n^3) / 31104$\\
\hline
2,3&$(-16860112 + 40609072\,m - 6888848\,m^2 - 29640128\,m\,n + 
   19136204\,m^2\,n - 465020\,m^3\,n - 6579128\,m^2\,n^2 $\\
&$+ 911110\,m^3\,n^2 - 
   210\,m^4\,n^2 - 223255\,m^3\,n^3 + 420\,m^4\,n^3 -105\,m^4\,n^4)/ 5971968$\\
&$+\pi^4(-14534 + 34544\,m-5476\,m^2-24459\,m\,n+14676\,m^2\,n - 302\,m^3\,n - 
   4923\,m^2\,n^2 $\\
&$+ 616\,m^3\,n^2 + 30\,m^4\,n^2 - 127\,m^3\,n^3 - 
   60\,m^4\,n^3 + 15\,m^4\,n^4) / 1866240$\\
&$+\pi^6(-10110 +24060\,m-3840\,m^2-16915\,m\,n + 9970\,m^2\,n - 200\,m^3\,n - 
   3265\,m^2\,n^2 + 400\,m^3\,n^2 - 100\,m^3\,n^3) / 5878656$\\
&$+\zeta(3)(-178724 + 417584\,m - 60136\,m^2 - 277386\,m\,n + 145188\,m^2\,n - 
   8000\,m^3\,n - 51406\,m^2\,n^2 + 17930\,m^3\,n^2 $\\
&$- 170\,m^4\,n^2 - 
   5135\,m^3\,n^3 + 340\,m^4\,n^3 - 85\,m^4\,n^4)/ 124416$\\
&$+\zeta^2(3)(6522-15252\,m+2208\,m^2+10313\,m\,n-5414\,m^2\,n + 40\,m^3\,n + 
   1643\,m^2\,n^2 - 80\,m^3\,n^2 + 20\,m^3\,n^3) / 15552$\\
&$+\zeta(5)(-1320 + 3480\,m -840\,m^2-2486\,m\,n+ 1376\,m^2\,n + 116\,m^3\,n - 
   89\,m^2\,n^2 - 358\,m^3\,n^2 + 121\,m^3\,n^3) / 31104$\\
\hline
1,4&$( 1 - 2 m + m n )(315(3166528+1077120\,m\,n + 45254\,m^2\,n^2 + 21\,m^3\,n^3)
+1008 \pi^4(2668+ 816\,m\,n + 29\,m^2\,n^2 - 3\,m^3\,n^3)$\\
&$+3200\pi^6(186 + 55\,m\,n + 2\,m^2\,n^2)  
+15120\zeta(3)(31848 + 8208\,m\,n + 940\,m^2\,n^2 + 17\,m^3\,n^3)$\\
&$-241920 \zeta^2(3)(582 + 145\,m\,n + 2\,m^2\,n^2)
+241920 \zeta(5)(72 + 14\,m\,n - 5\,m^2\,n^2))  /752467968$\\
\hline
0,5&$- ( 2 + m n ) (315(3166528+1077120\,m\,n + 45254\,m^2\,n^2 + 21\,m^3\,n^3)
+1008 \pi^4(2668+ 816\,m\,n + 29\,m^2\,n^2 - 3\,m^3\,n^3)$\\
&$+3200\pi^6(186 + 55\,m\,n + 2\,m^2\,n^2)  
+15120\zeta(3)(31848 + 8208\,m\,n + 940\,m^2\,n^2 + 17\,m^3\,n^3)$\\
&$-241920 \zeta^2(3)(582 + 145\,m\,n + 2\,m^2\,n^2)
+241920 \zeta(5)(72 + 14\,m\,n - 5\,m^2\,n^2))  / 3762339840$\\
\end{tabular}
\label{tabnu}
\end{table}

\end{document}